\documentclass{aa}  
\usepackage{hyperref}
\usepackage{placeins}

\usepackage{xcolor}
\hypersetup{    
    colorlinks=true,
    citecolor = blue,
    linkcolor=blue,
    filecolor=magenta,      
    urlcolor=cyan,
}
\usepackage{multirow}

\usepackage{graphicx}

\usepackage{txfonts}
 \def\mso{\,\mathrm{M}_\odot}

 \def\msoy{\, \mso~{\rm yr}^{-1}}
 \def\kms{\, \rm km\,s^{-1}}
 
 \def\Msun{\,\mathrm{M}_\odot}
 \def\Rsun{\,{\rm R}_\odot}

\newcommand{\case}[1]{Case\,#1}
\newcommand{\Case}[1]{Case\,#1}

 \def\qi{q_\mathrm i}
 
 \def\days{\,\text{d}}

 \def\Mej{M_\mathrm{ej}}
 \def\Mcsm{M_\mathrm{CSM}}
 \newcommand{\Type}[1]{\text{Type}\,\text{#1}}

 \newcommand{\multilinecomment}[1]{}

 \newcommand{\de}{\mathrm{d}}

\def\simle{\mathrel{\hbox{\rlap{\hbox{\lower4pt\hbox{$\sim$}}}\hbox{$<$}}}}
\def\simgr{\mathrel{\hbox{\rlap{\hbox{\lower4pt\hbox{$\sim$}}}\hbox{$>$}}}}

\defcitealias{Ercolino_widebinary_RSG}{EJLD24}
\defcitealias{Ercolino_HpoorInteractingSNe}{EJLD25}
\defcitealias{MHLC16}{M16}
\defcitealias{Ertl16_explodability_m4mu4}{E16}
\defcitealias{PS20_explodability_Mco_Xc}{PS20}
\def\PI{\citetalias{Ercolino_widebinary_RSG}}
\def\PII{\citetalias{Ercolino_HpoorInteractingSNe}}
\def\Muller{\citetalias{MHLC16}}
\def\Ertl{\citetalias{Ertl16_explodability_m4mu4}}
\def\PS{\citetalias{PS20_explodability_Mco_Xc}}

\begin{document}

   \title{The demographics of core-collapse supernovae}
   \subtitle{The role of binary evolution and CSM interaction}

   \author{Andrea Ercolino
          \inst{1}
          \and Harim Jin \inst{1}\and
          Norbert Langer\inst{1,2}  \and Avishay Gal-Yam \inst{3} \and Abel Schootemeijer \inst{1} \and Caroline Mannes\inst{1,2}
          }

   \institute{Argelander Institut für Astronomie,
              Auf dem Hügel 71, DE-53121 Bonn, Germany\\
              \email{aercolino@astro.uni-bonn.de}
         \and
          Max-Planck-Institut für Radioastronomie, Auf dem Hügel 69, DE-53121 Bonn, Germany
          \and 
Department of Particle Physics and Astrophysics, Weizmann Institute of Science, IL-76100 Rehovot, Israel}

   \date{Received Month Day, 2025; accepted Month Day, 2025}

  \abstract
{The observational properties of core-collapse supernovae are shaped by the envelopes of their progenitors. In massive binary systems, mass-transfer drastically alters the pre-supernova structures compared to single stars, leading to a diversity in supernova explosions. }
{We compute the distribution of core-collapse supernova properties based on comprehensive detailed grids of single and binary stellar evolution models.}
{We conduct a grid-based population synthesis to produce a synthetic population of core-collapse supernovae, and compare it to observed supernova samples. To do this, we apply various explodability and merger criteria to our models. In line with earlier results, we identify interacting supernova progenitors as those stars that undergo core-collapse during or shortly after a Roche-lobe overflow phase. }
{With an interacting binary fraction of 68\%, our models predict two-thirds of all core collapse supernovae to be of \Type IIP/L,
and one-third of \Type Ibc, in agreement with recent volume-limited supernova surveys. We find that 76\% of the \Type Ibc supernova progenitors took part in a previous binary mass transfer (mostly as mass donor), but also 63\% of the Type\,IIP/L supernova progenitors (mostly as mass gainers), yielding a much broader envelope mass distribution than expected from single stars.
We find that mass-transfer induced interacting supernovae make up $\sim5\%$ of all core-collapse supernovae, which is close to the observed fractions of \Type IIn and \Type Ibn supernovae. When assuming a disk or toroidal CSM geometry for \Type IIn supernovae, our models predict a bimodal distribution of the radiated energies similar to the distribution deduced from observations. 
}
{While we find the effect of binary evolution on the relative number of \Type Ibc and \Type IIP/L supernovae to be moderate,
it leads to lower average ejecta masses in \Type Ibc and \Type IIb supernovae, and can lead to higher pre-supernova masses in Type IIP/L supernovae than single stars. Binary models are also able to reproduce the number and properties of interacting supernovae. }

   \maketitle

\section{Introduction}

Core-collapse supernovae mark the end of the life of stars initially more massive than $M_\mathrm{i}\simgr9\Msun$ as their iron cores collapse and the released energy drives an explosion. These are among the most energetic events in the Universe, playing a crucial role in enriching their host galaxies with metals and triggering subsequent episodes of star formation \citep{McKeeOstriker_ISM_SNe, MacLow2004_supersonic_turbolence_star_formation_SNe, MacLow2005_pressures_ISM_SNe, Nomoto2006_nucleo_sne_galactic_chemical_evolution}.

Supernovae are classified observationally based on the presence of hydrogen in their spectra \citep[see][for a review]{GalYam17_SN_Book}. H-rich events are labeled \Type IIP or \Type IIL depending on the light-curve, while those whose progenitors were partially or fully stripped of the H-rich envelope give rise to \Type IIb and \Type Ibc supernovae, respectively \citep{Young2004_LC_parameters,Dessart2011_IIb, Dessart2024_partiallystripped_typeII_progs}.   All stripped-envelope supernovae provide a testbed for massive star evolution models. As single-star models under-predict their observed rates, several authors proposed binary evolution as a key formation channel \citep{Eldridge08_massivebinaries_sntypes,Smartt_rev_2009,Smith2011_obsSNfractions,Eldridge13_deathmassivestarII_Ibc}.

Binary mass transfer is understood to play a dominant role in shaping the evolution and final fate of massive stars \citep{Sana_massive_stars_binaries, Langer_review_2012}. Donor stars in close binaries can be efficiently stripped of most, if not all, of their H-rich envelopes, producing H-poor or even He-poor supernovae \citep[e.g.,][]{Podsiadlowski_massive_star_binary_interaction_1992,WL99_closebinaries_BH&SNe, Yoon2010_Ibc,   Yoon_IIb_Ib, Claeys_b, Sravan_b}. Mass transfer also alters the structure and explosion timing of mass donors and mass gainers compared to single stars \citep{Wellstein_2001_ContactMassive,Zapartas2017_companions_in_SESNe_Ic, Eldridge17_BPASS, Laplace_core_single_vs_binary_2021}.

Previously, the progenitors of \Type Ibc supernovae were considered to be massive H-free Wolf-Rayet stars (e.g., \citealt{Georgy09_progenitors_IbIcGRB, Groh13_singlestarevo}). The current consensus has instead shifted to include progenitors that underwent binary-induced mass-loss  \citep{WLW1995_preSN_HeS, WL99_closebinaries_BH&SNe, Yoon2010_Ibc, Yoon_IIb_Ib}, especially in light of transients with low-mass progenitors like \object{iPTF13bvn} \citep{Bersten2014_iPTF13bvn,Fremling14_iPTF13bvn_notWR,Kuncarayakti15_iPTF13bvn_nebular_binary, Yoon_IIb_Ib}.  

Some supernovae show strong observational signatures of interaction with the surrounding circumstellar medium (CSM), implying significant mass-loss in the final stages before explosion \citep{Smith17_InteractingSNe_book}. These transients are categorized based on the composition of the narrow emission lines: H-rich (\Type IIn, \citealt{Schlegel90_IIn}), H-poor and He-rich (\Type Ibn, \citealt{Pastorello08_2006jc_Ibn}) and H-poor and He-poor (\Type Icn, \citealt{GalYam2022_SN2019hgp_Icn}).   
Recent transient surveys (e.g., ATLAS, \citealt{ATLAS}, ASAS-SN, \citealt{ASASN}, PTF/iPTF, \citealt{PTF}, and ZTF, \citealt{ZTF}) have uncovered a larger number of such events which are substantially diverse \citep[e.g.,][]{Hiramatsu25_IIn_bimodality}.   
One of the main avenues of research is to connect the progenitor stars and their evolution with these observed supernovae.  

Models for interacting supernovae fall into two broad categories. One attributes the mass-loss to single-star related mechanisms that take place just before core-collapse, such as LBV-eruptions or instabilities in the envelope or core \citep{Woosley1980_TypeI_flashes, Quataert_Shiode_wavedriven_winds, SmithArnett2014_Hydroinstab_Turb_preSN, Woosley_Heger_2015_SiFlash, Fuller17_waveheating_RSG, WuFuller21_wavedrivenoutburst, Schneider25_accretorsmergerSNe_IIP_87A_IIn}. The alternative scenario involves inefficient mass-transfer in binary systems occurring in late evolutionary stages
(\citealt{Ouchi2017_IIb_RSG_progenitors, Matsuoka_Sawada_BinaryInteraction_IIP_Progenitors, Ercolino_widebinary_RSG} for the H-rich transients and \citealt{WuFuller_IbcLMT, Ercolino_HpoorInteractingSNe} for the H-poor ones).

In this paper, we present a grid-based binary population synthesis study informed by a comprehensive set of detailed stellar and binary evolution models \citep{Jin25_BonnGal} aimed at predicting the relative rates and progenitor properties of various core-collapse supernova types. Building on our earlier works on interacting supernovae (\citealt{Ercolino_widebinary_RSG} for \Type IIn supernovae, hereafter \PI, and  \citealt{Ercolino_HpoorInteractingSNe} for \Type Ibn supernovae hereafter \PII) we provide quantitative estimates of the populations of interacting supernovae.

We will introduce our models and assumptions in Sect.\,\ref{sec:methods}, and discuss our results in Sect.\,\ref{sec:results}. We will compare our results with the observations in Sect.\,\ref{sec:results:observations} before concluding in Sect.\,\ref{sec:conclusions}.

\FloatBarrier

\section{Method} \label{sec:methods}

\subsection{Stellar evolutionary models}\label{sec:methods:models}

\begin{table}
    \centering
    \caption{List of stellar evolutionary model grids used in this work.}
    \resizebox{\columnwidth}{!}{
\begin{tabular}{r|cccc}
   Grid & B & S & CC-S & CC-He \\\hline\hline
    System & Binary & Single & Single & Single HeS \\
    $M_\mathrm{i}$ or $M_\mathrm{1,i}$ [$\Msun$] & $5-100$ & $5-12.6$ & $12.6-200$ & $2.5-40.7$ \\
    $P_\mathrm{i}$ [$\days$] & $1.3-5011$ & - & - & - \\
    $\qi$ & $0.10-0.95$ & - & - & - \\
    $v_\mathrm{rot,i}$ & $0.2\,v_\mathrm{crit}$ & $150\kms$ & $0.2\,v_\mathrm{crit}$ & No \\
    Network & \multicolumn{2}{c}{\texttt{stern}} & \multicolumn{2}{c}{\texttt{approx21}}  \\
    Termination & \multicolumn{2}{c}{core C depletion} & \multicolumn{2}{c}{core-collapse} \\
    Reference & [1] & [2] & \multicolumn{2}{c}{This work} \\
\end{tabular}
}
    \tablefoot{The first row shows the acronym with which we refer to each grid. Each row then shows whether they evolve a binary-star system, a single-star system or a single  {helium star (HeS)} system (2), the initial mass  {of the stellar model,}  {or of the initially more massive star if it is in a binary system} (3), initial orbital period and mass-ratio (4-5),  rotation (6), the nuclear network adopted (7), the latest evolutionary phase that can be captured by these models (8), and the reference (9). \textbf{References}: [1] \citealt{Jin25_BonnGal}, [2] \citealt{Jin2024_boron}.}
    \label{tab:grids}
\end{table}

We consider stellar evolution models from the Bonn-GAL grid \citep{Jin25_BonnGal} computed with the code MESA \citep[r10389,][]{MESA_I, MESA_II, MESA_III, MESA_IV} that include differentially rotating single- and binary-evolution models at solar metallicity. Since these models are only evolved until the end of core C-burning, we additionally compute two single-star grids that reach core-collapse, enabling us to estimate explosion properties for progenitors across the parameter space (see Sect\,\ref{sec:methods:explodability}). The details of the grids can be found in Table\,\ref{tab:grids}. 

In the following, we discuss only the relevant physics and assumptions in the context of this work. We refer to \cite{Jin2024_boron} and \cite{Jin25_BonnGal}, for a complete description of the stellar evolutionary models.  A discussion of the effects of the physics used in the stellar models on our results can be found in Appendix\,\ref{sec:discussion:model_physics}. 

\subsubsection{Stellar physics}\label{sec:method:stellar_physics}
Stars are initialized with an initial rotational velocity of roughly $20\%$ of the breakup rotation \citep[corresponding to the first peak in the observed distribution in][roughly $150\kms$]{VLT_FLAMES_X}. 
The Bonn-GAL grid adopted the \texttt{stern} nuclear network \citep[][]{HLW2000, Jin2024_boron, Jin25_BonnGal} and the simulations are terminated at the depletion of C in the core. 
Lower-mass models are terminated when they develop into asymptotic giant branch stars.

The models use the wind mass-loss recipe of \citet{Pauli_LMC} which is built upon that of \cite{Brott2011}. For hot H-rich stars ($X>0.7$), the wind mass-loss rate is taken from \citet{Vink2001}, while for cool H-rich stars the maximum between the rates of \citet{Vink2001} and \citet{Nieuw_wind} is adopted. For H-poor stars ($X<0.4$), the Wolf-Rayet wind mass-loss rates from \citet{NugisLamers2000} are adopted, until the star becomes H-free, at which point the rates from \citet{Yoon_wind} are implemented.
Rotationally enhanced mass-loss is also accounted for following \cite{HLW2000} which results in a boost of a factor $(1-v_\mathrm{rot}/v_\mathrm{crit})^{-0.43}$, where the critical velocity is also a function of the Eddington factor.

Convective regions are treated via the mixing-length theory with $\alpha_\mathrm{MLT}=1.5$.  Stellar models show subsurface convectively-unstable regions where the envelope is locally super-Eddington, and result in inflated envelopes \citep{Sanyal2015_supereddington_envelopes} which give rise to numerical instabilities in extended stars with high luminosity-to-mass ratio $L/M$. Therefore the Bonn-GAL models additionally include MLT++ \citep{MESA_II} after the end of the main-sequence to reduce the superadiabaticity of the outermost layers. This treatment also results in systematically smaller radii in affected stars, thus influencing the onset of mass-transfer episodes in partially-stripped stars (see Appendix\,\ref{sec:discussion:singlestar}). The convective boundaries are determined via the Ledoux criterion, and semiconvection is included with an efficiency parameter $\alpha_\mathrm{sc}=1$ \citep{Schootemeijer2019_mixinginmassivestarsSMC}.

\subsubsection{Binary physics}\label{sec:methods:binary_physics}
Binary models are initialized in a circular orbit, and the two stars are computed simultaneously until one of the two reaches an advanced evolutionary stage \citep{Jin25_BonnGal}. If one star is expected to reach core collapse, the binary is broken up and the companion's computation is continued until it also reaches the end of its life. If it is otherwise expected to become a white dwarf (WD), then it is set to a point mass, and the companion is seamlessly evolved in the binary. If there are convergence issues, the binary simulation is terminated. 

When stars evolve in binaries, they can undergo one or more phases of mass transfer via Roche-lobe overflow (RLOF), which is implemented via the ``contact'' scheme during the main-sequence \citep{MESA_II} and \cite{Kolb_scheme} after one star evolves out of the main-sequence. We assume rotation-limited accretion on the companion star, and that the material lost carries away the same specific angular momentum of the accretor. This results in the accretors not gaining significant amounts of mass, unless they are in tight binaries \citep{Jin25_BonnGal}.

Mass transfer can turn unstable, potentially leading the system into a common-envelope phase \citep[][]{Paczynski_tau_inspiral_CEE} where the companion star in-spirals within the donor star's envelope. We assume that this phase results in the merger of the two stars. We make an exception for unstable \Case C RLOF, as the in-spiral timescale may be comparable, or even exceed, the remaining time before core collapse allowing the donor to explode before the merger occurs (see \PI). 

Several criteria are used to identify binary models that undergo unstable mass transfer and function as termination conditions \citep{Jin25_BonnGal}, encompassing L2-overflow during the overcontact phase, mass transfer initiated at the Zero-age main-sequence (ZAMS), the occurrence of Darwin instability \citep{Darwin1879_StabilityBinary}, inverse mass transfer, and models with $\qi<0.7$ where mass transfer rates exceed $0.1\msoy$. The models that prematurely terminated during \Case A or \Case B mass transfer are assumed to merge. This ensemble of choices defines the ``Hardcoded'' merger criterion, outlining the baseline for the number of systems that will result in mergers.

\subsection{Population synthesis}\label{sec:method:popsynth}

We estimate the population properties of supernova progenitors using a grid-based population synthesis of the stellar evolutionary models, assigning statistical weights to each model in the grid. This is motivated by the widespread presence of sharp transitions in the parameter space, which disfavor the adoption of interpolation-based Monte Carlo methods.  

We describe our assumptions on the birth probabilities of each model (Sect.\,\ref{sec:methods:stat_weights}), and on the inferred properties of the models beyond the stellar-evolution calculations, namely the evolution of mergers (Sect.\,\ref{sec:methods:megers}), the explodability of the models (Sect.\,\ref{sec:methods:explodability}),  and the evolution of the binary system after the first explosion (Sect.\,\ref{sec:method:postSN}). We discuss the effect of these assumptions in Appendix\,\ref{sec:discussion:postprocess}.

\subsubsection{Birth probabilities}\label{sec:methods:stat_weights}

The single-star and binary grids cover an extensive parameter space for the initial mass $M_\mathrm{1,i}$, period $P_\mathrm{i}$ and mass-ratios $\qi$ (see Table\,\ref{tab:grids}). We adopt a power-law distribution for $M_\mathrm{i}$ (i.e. the initial-mass function, IMF, \citealt{Salpeter_IMF_55, Kroupa2001_IMF}) as well as for $\qi$ and $P_\mathrm{i}$, and we can write the overall probability density function of a system being born as a binary ($\de n_B$) or as a single star ($\de n_S$) as
\begin{equation}\label{eq:distribution}
\left\{    
\begin{aligned}
    \de n_B &\propto f_\mathrm B\cdot  M^{\alpha} (\log P)  ^{\beta}  q^{\gamma}   \ \de M \de  (\log  P )\de q \\
    \de n_S &\propto f_\mathrm S \cdot  M ^{\alpha}  \de M
\end{aligned},
\right.
\end{equation}
with $f_\mathrm{B}$ the birth-binary fraction, and $f_\mathrm{S}=1-f_\mathrm{B}$ the birth single-star fraction. The power-law exponents are taken as $\alpha=-2.35$ (\citealt{Salpeter_IMF_55}; note that for binaries we use the initial mass of the more massive component), while the most recent literature suggests $\beta\in[-0.1,-0.2]$ \citep{Almeida17_FLAMES_OB_binaries, Sana2025_BLOEM_binaryfraction_Z} and $\gamma=-0.10\pm0.58$ \citep{Sana_massive_stars_binaries}. Considering the relatively high uncertainty of these last two indices, we adopt a flat distribution with $\gamma=\beta=0$. In Appendix\,\ref{sec:discussion:distributions}, we elaborate on the effect that employing different distribution powers has on our results.

Each binary model in the grid (with initial conditions $\log M_\mathrm{1,i}$, $\log P_\mathrm{i}$, $\qi$) represents the systems within a small interval from its initial conditions. We take this interval as $\delta/2$ with $\delta=0.05$ being the grid-resolution in $\log M_\mathrm{1,i}$, $\log P_\mathrm{i}$ and $\qi$. Statistical weights are assigned by integrating the probability density (Eq.\,\ref{eq:distribution}) over the discrete volume elements. For single stars, we proceed analogously (where only the initial mass changes between models) while including a finer grid in mass by interpolating by the models' initial mass.

We sample our single-star and binary models assuming a mass-independent birth binary fraction of  {stellar systems of} $f_B=75\%$ (\citealt{Sana_massive_stars_binaries}\footnote{$71\%$ of  {galactic O-stars are observed with a binary companion in an orbital configuration such as to be expected to undergo significant mass-transfer in the future} before the first supernova (with $P_\mathrm{i}<1\,500\days$). This number extends to $75\%$ if one includes binaries with very wide orbits in which RLOF may not affect the two stars significantly, if it even occurs at all ($P_\mathrm{i}<3\,000\days$). Since our binary models' initial orbital period also include periods up to $5\,000\days$, we consider the binary fraction to be $75\%$ when weighting our binary models.}). 
 {Translating these to numbers of stars yields that} about $14.3\%$ of them are born as single-stars and $85.7\%$ are in a binary system. Our widest binary models do not undergo any mass-transfer, and we therefore define an empirical interacting-binary fraction $f_\mathrm{B}^\mathrm{MT}$ as the fraction of systems at birth that undergo mass-transfer. Assuming a Salpeter-IMF, flat-$\log P_\mathrm{i}$ and flat-$\qi$ distributions, we obtain $f_\mathrm{B}^\mathrm{MT}=67.8\%$, while for effectively-single star systems (which include wide binaries that do not undergo mass-transfer) the fraction is $f_\mathrm{S}^\mathrm{eff} = f_\mathrm{S}+(f_\mathrm{B}-f_\mathrm{B}^\mathrm{MT})=32.2\%$. This translates to $22.5\%$ of stars being born in single- or effectively-single star systems, while $77.5\%$ of stars undergo mass-transfer with a companion.

\subsubsection{Mergers and their evolution}\label{sec:methods:megers}
The evolution of a binary system after the onset of unstable mass-transfer is the topic of many recent studies \citep{Ivanova_book, SchneiderLauReopke_CEE_book} with focus on both multi-dimensional simulations \citep{Lau_Hydro_12M,  Lau25_12Msun_CEE, Gagnier_post_plunge_in, GagnierPejcha25_CEE} and one-dimensional approximations \citep{ Pablo_Kolb,Hirai_CE_2step, Bronner24_CEE}. While there is still significant debate over the physics of the common-envelope phase, these studies suggest that most of the systems like those which we label as undergoing unstable mass-transfer will merge.

We assume that unstable \Case A or \Case B mass-transfer between two non-degenerate stars always leads to a merger (see Sect.\,\ref{sec:method:postSN} for the case with a degenerate companion). The evolution of the merger product has been investigated in previous studies \citep{MenonHeger17_mergers, Menon24_mergermodels, Schneider2024_preSNevo_mergers_secondaries}, but it depends on the different physical assumptions adopted. Here, we simplify the treatment of these stars and separate them on whether they formed during \Case A RLOF, \Case B RLOF or inverse mass-transfer. 

For \Case A merger products, we assume complete rejuvenation such that the merger product truly resembles a single star of the same total mass \citep[which is justified considering the high semiconvection adopted,][]{Braun_Langer_95}. Its evolution is therefore mapped to the single-star model with the same mass. These stars may however show strong magnetic fields \citep{Schneider19_mergers_magneticF_nature} which are not included in the calculations. We assume that mergers where at least one star has developed a H-free core (i.e. \Case B and inverse mass-transfer) form a star where the H-free core and the envelope are the sum of those from the two progenitor stars. The final properties of these models at core collapse are evaluated by mapping our merger models (defined by their initial mass and the relative amount of mass accreted) to the models studied in \cite{Schneider2024_preSNevo_mergers_secondaries}. 

We explore additional criteria to identify models undergoing unstable mass-transfer and then merge (as compared to the ``Hardcoded'' criteria in Sect.\,\ref{sec:methods:binary_physics}) by post-processing the models in the grid. This allows us to discuss the effects these criteria have on the population of supernovae. We include two broad categories of merger criteria, which either define mass-transfer stability via the successful ejection of unaccreted material via radiation \citep[which we will refer to as the ``Energy'' criterion, ][]{PABLO_LMC_GRID} or the presence of outflows from the outer Lagrangian point (which we refer to as the ``OLOF'' criterion, \citealt{Pavlovskii_Ivanova_2015_MT_from_Giants}). These two criteria flag a model as undergoing unstable mass-transfer even if the respective condition is met for a short time. We therefore include a ``Delayed Energy'' criterion \citep{Pauli_MasterThesis} and a ``Delayed OLOF'' criterion (\PI) to only consider those that meet these conditions for longer timescales. 

\subsubsection{Explodability of the models}\label{sec:methods:explodability}
As models in the Bonn-GAL grids are not computed to core-collapse (see Sect.\,\ref{sec:method:stellar_physics}), we assume that the models' core and envelope masses after core He-depletion, when available, are representatives of those at core-collapse (unless mass transfer happens in between; see Sect.\,\ref{sec:methods:sn_type}). We assume that models successfully reach core-collapse when $M_\mathrm{CO-core}>1.43\Msun$ \citep{Tauris2015_USSNe}, $M_\mathrm{He-core}>2.49\Msun$ or $M_\mathrm{conv,He}^\mathrm{max} >1.30\Msun$ (the extent of the convective He-burning core, see \PII). Models that do not meet any of these conditions are instead assumed to produce WDs. 

To distinguish between failed and successful supernovae, we require knowledge on a star's structure at core-collapse ( {e.g.,} \citealt{ OconnorOtt11_compactness_and_bh_formation,MHLC16,Ertl16_explodability_m4mu4}, with a few notable exceptions such as \citealt{PS20_explodability_Mco_Xc} and \citealt{Maltsev2025_explodabilityML}). To remedy this, we map the models in the Bonn-GAL grid to the appropriate core-collapsing models (see Table\,\ref{tab:grids}) with the closest CO-core mass (or He-core mass if the former is not available). 

We computed two grids of single-star models to core-collapse. The first grid (Grid CC-S) includes stars computed from ZAMS (with the same physics as \citealt{Jin2024_boron}) and is used to infer the explodability of effectively single-stars or those that have undergone \Case C RLOF \citep{Schneider_preSNevo_stripped_2021}. We also map \Case A mergers (see Sect.\,\ref{sec:methods:megers}) and accretors to this grid.
For \Case B mergers, we map the models to those from Grid CC-S with core-masses that are $0.5\Msun$ higher \citep{Schneider2024_preSNevo_mergers_secondaries}.   The second grid of core-collapsing models is built to mimic binary-stripped He-stars (Grid CC-He, constructed as in \citealt{DR1, DR2}), and models that were stripped during \case A or \Case B RLOF are mapped to this grid \citep{Schneider_preSNevo_stripped_2021}. Models with small core-masses could not be computed successfully to core-collapse. Therefore, models that are expected to reach core-collapse with small core-masses, are always mapped to the least-massive model in either CC-S or CC-He (i.e. $M_\mathrm{ZAMS}=12.6\Msun$ or $M_\mathrm{HeS}=4.0\Msun$ respectively).

 To assess the explodability of the models at core-collapse, we employ different methods and criteria.  We will use the semi-analytical method from \cite{MHLC16}, hereafter \Muller, using the parameter calibration of \citealt{Schneider_preSNevo_stripped_2021} (other calibrations, including from \Muller \ and \citealt{DR2}, do not significantly alter the explodability of the models), which also provides estimates for the mass of the remnant and the explosion energy. We note that they will be typically overestimated in the population as the low-mass exploding models are mapped to higher-mass models that we were able to compute to core collapse.  
 
 We also use the criterion of \citet[][with the calibration N20 from \citealt{Ertl2020_explodability_HeS}]{Ertl16_explodability_m4mu4}, hereafter \Ertl. We also consider the explodability landscape shown in \cite{PS20_explodability_Mco_Xc}, hereafter \PS,
 which associates the explodability of a model from its properties at an earlier evolutionary stage (such as the CO-core mass $M_\mathrm{CO}$, and the central abundance of carbon at core He depletion $X_\mathrm{C}$; cf. \citealt{Brown2001_HMXB_BH_formation}). Since their models cover $M_\mathrm{CO}\leq10\Msun$, we assume implosions for higher masses. 
 This method, while building on the method from \Ertl, allows us to map the explodability of any model without needing to calculate them until core collapse. The effect of different explodability assumptions on the single star grids CC-S and CC-He is discussed in Appendix\,\ref{appendix:single_star}.

 {It is important to note that all the methods and criteria listed here are all based on one-dimensional calculations or parameterizations, while it is now clear that multi-dimensional processes are key to understanding whether or not the shock following core-collapse successfully leads to an explosion \citep[e.g., ][]{BurrowsVartanyan21_review_CCSNe, Janka2025_review}}.  {Additionally, even though} we adopt the all-or-nothing approach for the formation of black holes (BHs), there may be cases where the fallback following a successful explosion \citep[e.g.,][]{Colgate1971_NSformation_fallback, Chevalier1989_NS_accretion_fallback, WoosleyWeaver95_explosionsofmassivestars,Fryer1999_fallback_bh_formation,Zhang2008_fallback_and_bhs,Chan18_fallback40Msun_Arepo, Burrows25_BH_formation_CCSN} may lead to the formation of a BH. This implies that the ejecta masses derived in this work are to be taken as upper limits. At the same time, we would form more BHs, which would affect the mass distribution we derive.

\subsubsection{Evolution of the companion following the first supernova}\label{sec:method:postSN}
For successful explosions, we assume that the natal kick on the newly-born neutron star (NS) is enough to break-up the binary, as expected in the vast majority of cases \citep{DeDonder97_Runaways_kicks, BrandPodsi95_SNkicks_and_NSbinaries, Eldrige11_runaway_progenitors_SNe, Renzo19_runaway_postSNbreakup, Xu25_SMC_pop_mesa, Schuermann25_SMC_pop_combine}. This simplified treatment will place an upper limit on the contribution of secondary stars to the population of supernovae (see the discussion in Appendix\,\ref{sec:discussion:postSN_mergers}) . 

For systems in which one star forms a BH, we assume no kick is imparted, and we extrapolate the evolution of the companion based on its radius evolution and orbital separation. If the star would fill its Roche-lobe during the main-sequence, we assume the mass transfer to turn unstable, and the star is disrupted. If mass transfer instead occurs after the main-sequence, we assume the mass transfer phase to strip the donor star of its H-rich envelope. For those systems that undergo RLOF after core He-depletion, we assume that some mass is lost from the system forming the CSM which will turn the resulting supernova (if the star successfully explodes) into an interacting supernova (Sect.\,\ref{sec:method:interaction}). 

These assumptions on mass-transfer are rough approximations, and recent works suggest that some stars undergoing \Case A mass transfer with a BH may actually survive while some undergoing \case B would merge \citep{Klencki25_OB_BH_binaries_and_BBH_mergers}. We will discuss the effects of these assumptions in Appendix\,\ref{sec:discussion:postSN_mergers}.

\subsection{Supernova classification}\label{sec:methods:sn_type}

\subsubsection{Type IIP/L, IIb and Ibc supernovae}\label{sec:methods:sn_type:canonicalSN}
We define \Type IIP/L  supernovae as those in which the progenitor is H-rich ($M_\mathrm{H,env}\geq1\Msun$, with $M_\mathrm{H,env}$ the mass of the layers that contain H), but we will refrain from distinguishing between subcategories such as Type IIP and IIL which can be a function of mass in the envelope \citep{ Dessart2024_partiallystripped_typeII_progs} but also of pulsations \citep{Bronner25_IIP_IIL_Pulsation, Laplace25_pulsations_RSGs_2023ixf}. Stars that retain only a low-mass envelope at core-collapse ($0.001\Msun<M_\mathrm{H,env}<1\Msun$) are assumed to develop into a \Type IIb supernova (\citealt{Dessart2011_IIb}) while models with even lower H-rich envelope-masses are assumed to develop into a \Type Ibc supernova. See Appendix\,\ref{appendix:single_star} for the expected supernovae from single-star evolution. 

\Type Ibc supernovae can be further distinguished between \Type Ib and \Type Ic supernovae based on the amount of metals and He in the ejecta, but there is still uncertainty when connecting stellar progenitors to these subtypes, as it depends not only on the amount of He in the envelope, but also Ni-mixing during the explosion \citep{Dessart2012_IbcSNe, Hachinger2012_hidden_H_He_in_Ibc, Dessart2015_IIb_Ibc_RadTransf, Dessart2016_IIb_Ibc_RadTransf2, Dessart2020_Ibc, Teffs2020_H_He_hidden_in_Ibc_SNe,Williamson21_1994I_Ic_with_He}. Furthermore, photometric observations cannot distinguish these two classes (e.g., \citealt{Drout11_systematic_obs_study_Ibc}, but see \citealt{Jin2023_Ibc_color}) and spectra are always necessary to disentangle the two. We therefore refrain from making predictions of the two distinct classes. 

\subsubsection{SN1987A-like supernovae}\label{sec:methods:sn_type:87A}

Some stars may appear as a blue supergiant (BSG) prior to the explosion, which we assume occurs when $\log T_\mathrm{eff}/\mathrm{K}>3.9$ \citep{Schneider2024_preSNevo_mergers_secondaries}. This can be due to an over-massive H-rich envelope \citep[e.g.][]{Helling83_accretion_BSG, Podsi90_87A_merger,Podsiadlowski_massive_star_binary_interaction_1992, Justham2014_LBV_SLSNe_Mergers_BSG, MenonHeger17_mergers, Schneider2024_preSNevo_mergers_secondaries} as well as a partially-stripped envelope (\citealt{WoosleyPintoWeaver1987_progenitor87A_partiallystripped}, \PI). In some instances the star may remain as a BSG until core-collapse and therefore affect the resulting supernova, as was the case with \object{SN1987A} \citep{Walborn87_BSG_progenitor_1987A,WoosleyPintoWeaver1988_progenitor87A, Braun_Langer_95}.  

The H-rich supernova progenitors that are not merger products appear as BSGs at core C depletion only if they have a small H-rich envelope ($M_\mathrm{H,env}<1\Msun$, see also \citealt{Jin25_BonnGal}). The use of MLT++ is also expected to affect partially-stripped models by making them more compact and hence appear bluer (Appendix\,\ref{sec:discussion:singlestar}). We therefore ignore partially-stripped models when describing the  supernovae whose progenitors were BSGs, and only consider mergers. Since we do not follow the evolution of merger products, we compare our merging binaries to the models shown in \cite{Schneider2024_preSNevo_mergers_secondaries} with the same initial mass and accreted fraction to determine which are expected to appear as BSGs at core-collapse. We label these H-rich supernovae coming from BSG progenitors as 87A-like supernovae.

\subsubsection{Interaction with circumstellar medium}\label{sec:method:interaction}
We assume that binary systems undergoing mass transfer (either stable or unstable) after core He depletion of the supernova progenitor will develop into an interacting supernova (see \PI \ and \PII). While the presence of the CSM alone is not enough to guarantee whether the supernova will develop observable features typically associated with CSM-interaction, this will provide an upper-limit estimate on the number of interacting supernova we can produce. 

We classify those supernovae whose H-rich progenitors explode following stable or unstable \case C RLOF where a H-rich CSM has been produced as \Type IIn supernovae (\PI). We conversely identify \Type Ibn supernovae as those produced by a stripped He star after producing a He-rich CSM by \Case BB RLOF (see \PII). As the systems undergoing \Case BB RLOF are known to be underrepresented in the Bonn-GAL grid, we use the method illustrated in \PII\, to characterize the phase of \case BB RLOF that is otherwise missing in the models. We also provide a fitting a function to derive the amount of mass transferred during \Case BB RLOF based on properties of the stripped He stars and the orbit at core He depletion (see Appendix\,\ref{appendix:CaseBB}).

Assuming inelastic collision between the ejecta and a slow-moving CSM, we can measure the conversion rate of the explosion energy $E_\mathrm{kin,ej}$ to radiation $E_\mathrm{rad}$ via CSM-interaction using the mass of the ejecta and that of the CSM, which is given by (\citealt{Moryia13_CSMinteraction_2006gy, DRAD18_GRB_SLSN_Ic}, \PI, and \PII) as
\begin{equation}\label{eq:fM}
    f_M(\theta) = \sin\theta \frac{\Mcsm}{\Mcsm + \Mej\sin\theta},
\end{equation}
where $\theta$ is the half-opening angle from the orbital plane where the CSM is located (such that $\theta=90^\circ$ corresponds to the spherically symmetric case).

\section{Supernova population syntheses}\label{sec:results}

Here we describe the results of our population synthesis calculation. We derive the relative numbers of supernovae of different types as defined in Sect.\,\ref{sec:methods:sn_type}, as well as the distribution functions of their key properties. We start with our fiducial model in Sect.\ref{sec:results:example}, and then explore the major uncertainties affecting our models, i.e., the criteria used to assess their explodability in Sect.\,\ref{sec:results:exp}. We emphasize that we do not consider our fiducial model as more likely than the alternative ones discussed in Sects.\,\ref{sec:results:exp} and Appendix\,\ref{sec:results:parameters} (where we vary the merger criterion). We provide a brief overview of the effects that mass-transfer has on the supernova of otherwise single stars in Appendix\,\ref{sec:SNprog:Binary_Grid}.

\subsection{Fiducial population model}\label{sec:results:example}

\begin{figure}
    \centering
    \includegraphics[width=1\linewidth]{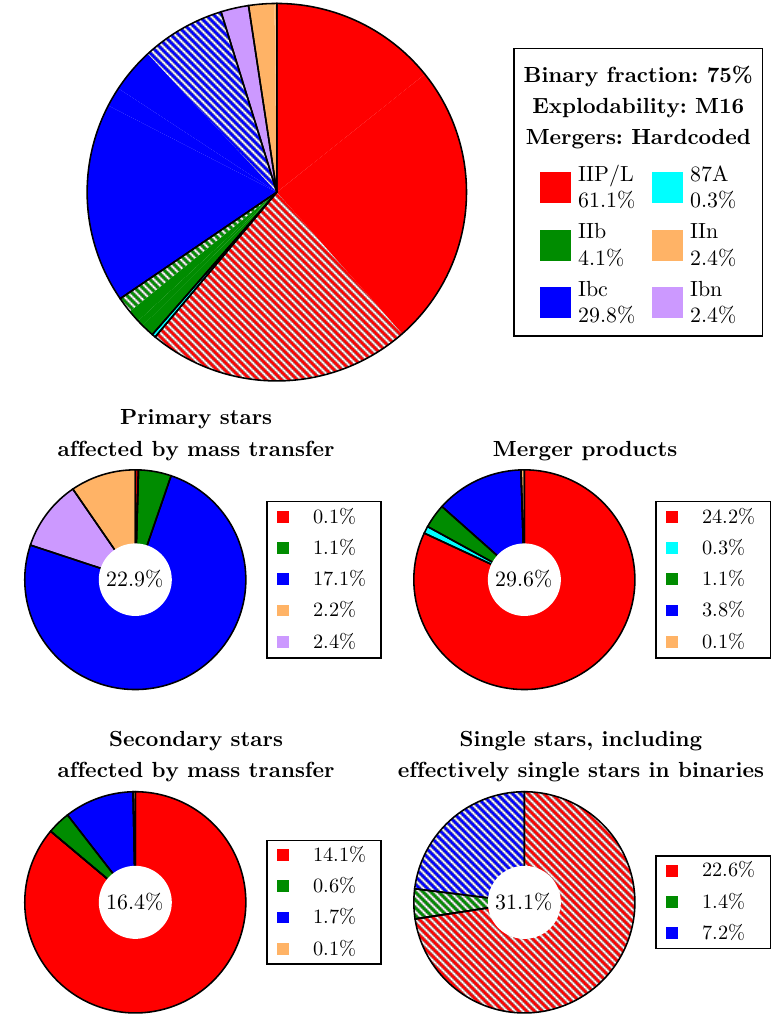}
    \caption{Pie-charts of the distribution of supernovae types (different colors) from the models assuming the explodability criterion of \Muller \ and the ``Hardcoded'' merger criteria. The contribution from stars that do not undergo mass-transfer is marked with a white hatching. The four lower panels represent the contribution from each progenitor type (primary and secondary stars that were affected by mass-transfer, merger products and single and effectively-single stars), with the contribution of that channel given in the center of the pie. The legends show the contribution to each supernova type to the whole distribution. We show a comparable plot based on the explodability criteria of PS20 in Fig.\,\ref{fig:overview_sne_PS20a}}
    \label{fig:overview_sne}
\end{figure}

Here, we calculate the population of supernovae assuming the explosion criteria of \Muller  \ and the ``Hardcoded'' merger criterion. As shown in Fig.\,\ref{fig:overview_sne}, this model predicts that $\sim 61\%$ of all supernovae are H-rich \Type IIP/L supernovae, $29\%$ are classical \Type Ibc supernovae. The remaining ~10\% are composed of Types\,IIn (2.4\%), IIb ($4.3\%$), Ibn (2.4\%) and\,87A (0.3\%). These numbers are qualitatively in line with previous theoretical works, which we discuss in Appendix\,\ref{sec:disussion:other_works}.

We find that about a third of all supernovae are produced by effectively single stars, i.e., stars which never participated in binary mass transfer. The progenitors of about two-thirds of all supernovae are binary interaction products, with merger stars producing $\sim30\%$ of all supernovae, and mass donors 23\%. Secondary stars which took part in at least one mass transfer event contribute less ($\sim$16\%), because  {they} are the initially less massive component in the binaries and therefore often form WDs. 

Primary stars, which are typically mass-donors, produce a population of supernovae that differs most strongly from that of single stars. Their majority explode as a \Type Ibc supernova, while in particular the long-period binaries with the lowest initial primary masses are not fully stripped and produce \Type IIb supernovae (see Fig.\ref{fig:logPq_1.10}). Per definition, the mass donors which reach core collapse during  mass transfer  are producing the interacting supernovae (Types\,IIn and\,Ibn). 

Secondary stars, merger products, and single stars produce mostly \Type IIP/L supernovae.
The most massive of them are able to self-strip via winds and also explode, which has a strong impact on the ejecta mass distribution of this class (see below).
Notably, this contribution is much smaller when the explodability criterion of \PS \ is applied (Sect.\,\ref{sec:results:exp}).

\begin{figure}
    \centering
    \includegraphics[width=1\linewidth]{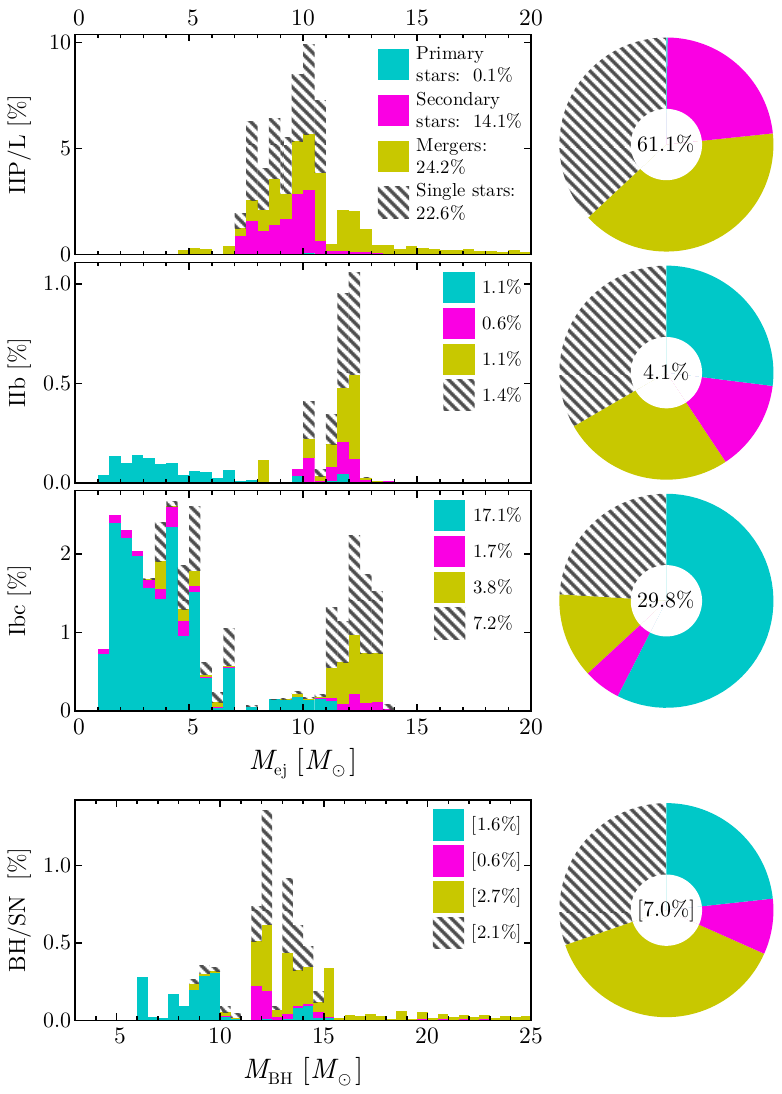}
    \caption{Stacked histograms of the ejecta mass $M_\mathrm{ej}$ in our fiducial population model grouped by the supernova type (\Type IIP/L,  IIb, and \Type Ibc supernovae) and normalized by the total number of supernovae (in percent). The contribution is split between primary stars (teal), secondary stars (magenta), merger products (dark gold) or single and effectively single stars (gray). The contribution from stars that have not undergone mass transfer is hatched. The pie-charts show the relative contribution of different progenitors to each supernova type. The legend also shows the fraction of supernovae within the whole distribution, and their sum is reported inside the pie-charts. Separately, we show a histogram for the mass of BHs with a pie-chart showing the relative contribution of different progenitors, and the number is normalized to that of supernovae. Less than one percentile of the distribution of \Type IIP/L supernovae and BHs fall outside of the shown ranges, extending up to $43\Msun$. We show a comparable plot based on the explodability criteria of PS20 in Fig.\,\ref{fig:overview_sne_PS20b}}
    \label{fig:overview_sne_bin_typeb}
\end{figure}

\subsubsection{Type IIP/L supernovae}
Roughly $\sim63\%$ of the \Type IIP/L supernovae in our synthetic population arise from mass gainer ($23\%$) or mergers ($40\%$) in binary systems (Fig.\,\ref{fig:overview_sne_bin_typeb}). Overall, these supernovae have ejecta masses of\footnote{All ranges of values provided in this work correspond to the interval between the 5th and 95th percentile of the derived distribution.} $M_\mathrm{ej}=7.5-14.4\Msun$, with a pre-explosion envelope mass of $M_\mathrm{H,env}=2.8-11.3\Msun$ (with half of them focused around $6.3-7.0\Msun$). These distributions show higher masses and are broader than those expected from single star evolution (see Appendix\,\ref{appendix:single_star}, $M_\mathrm{ej}=7.6-10.7\Msun$ and $M_\mathrm{H,env}=2.8-7.0\Msun$).  {The absence of more lower-mass envelopes is due to both our definition of \Type IIb progenitors and the fact that RSGs undergoing case C RLOF typically produce partially stripped enveloped (\PI, \citealt{Fang25_RSGs_diversity_Menv}), but are classified here as Type IIn progenitors rather than \Type IIP/L.}

While the bulk of secondary-star progenitors of \Type IIP/L supernovae have similar ejecta masses to those expected from single-star evolution, they also show a weak high-mass tail from $10.8$ to $24.4\Msun$ (which contribute to about $2\%$ of all \Type IIP/L supernovae). This high-mass tail is produced in \Case A systems where secondaries accrete significant amounts of mass (as tides spin them down during mass-transfer) and do not rejuvenate.  

Merger products have $M_\mathrm{ej}=7.1-17.8\Msun$ and $M_\mathrm{H,env}=3.1-13.8\Msun$. There are a few low-mass outliers ($M_\mathrm{ej}<7.4\Msun$, making up $\simle 2\%$ of all \Type IIP/L supernovae). These are the products of those mergers where the preceding phase of mass-transfer shed more mass than the mass the secondary star will contribute to. Most remarkable is the massive end of the distribution ($M_\mathrm{ej}>10.8\Msun$, comprising about $14\%$ of all \Type IIP/L supernovae). This is produced by \Case B mergers, where the addition of the secondary star's mass produces an over-massive envelope. Progenitors with such massive hydrogen envelopes are not produced by single stars, and are a robust prediction of our binary models, independent of the adopted merger or explodability criteria.  

About $18\%$ of the merger progenitors of \Type IIP/L supernovae (corresponding to about $7\%$ of all \Type IIP/L supernovae) come from \Case A system with $M_\mathrm{1,i}<10\Msun$. This is possible as the two stars, which initially were not massive enough to have reached core-collapse, merge to form a more massive star that is able to explode.

\subsubsection{Type IIb supernovae}

About two thirds of \Type IIb progenitors are produced in mass-transferring binaries ($\sim 68\%$, Fig.\,\ref{fig:overview_sne_bin_typeb}).  Primaries contribute significantly to this population ($27\%$), and always arise from systems that have undergone late \Case B mass transfer (see Fig.\,\ref{fig:logPq_1.10}). This is in contrast with the discussion on the nature of the progenitor of the prototype \Type IIb \object{SN1993J} (e.g., \citealt{1993J_Podsiadlowski}), which proposed that the progenitor was produced via \Case C RLOF. Notably, our models predict that such Case\,C  progenitors would instead give rise to interacting Type\,II supernovae (see Sects.\,\ref{sec:methods:sn_type}). Our models suggest that even at high metallicity, primary stars that underwent \Case B RLOF can still retain sufficient hydrogen to appear as a \Type IIb supernova.  

Secondary stars, merger products, and single stars become \Type IIb supernovae only when they are massive enough to self-strip via winds. Such models produce very massive Type\,IIb progenitors, with ejecta masses between 8 and 13$\mso$, which make up $\sim 73$\% of them (Fig.\,\ref{fig:overview_sne_bin_typeb}). We show below that this very massive Type\,IIb supernova population disappears when the \PS \ explodability criterion is used, instead of that from \Muller \ (see Fig.\,\ref{fig:overview_sne_PS20b}).

\subsubsection{Type Ibc supernovae}

Three quarters of \Type Ibc supernovae are produced in mass-transferring binaries (Fig.\,\ref{fig:overview_sne}), and we observe a bimodal distribution in $M_\mathrm{ej}$ much like with the \Type IIb progenitors, with $69\%$ of the progenitors located in a broad lower-mass peak ($M_\mathrm{ej}=1.7-6.4\Msun$), and the remaining $31\%$ in a higher-mass peak ($M_\mathrm{ej}=9.5-13.4\Msun$). 

The first peak is populated mostly by primary stars that have been fully stripped following RLOF. The secondary stars that contribute to this group are stripped during mass-transfer with a BH companion. For $3.5\Msun\leq\Mej\leq7.0\Msun$ we find a contribution from very massive stars (including those that have not undergone any mass-transfer) which end their lives as low-mass He-stars at core-collapse (see Appendix\,\ref{appendix:single_star}). The higher-mass peak centered around $\sim 12\mso$ is populated mostly by effectively single stars and merger products. As for the Type\,IIb progenitors, this high mass peak disappears with the use of the \PS \ explodability criterion (see Fig.\,\ref{fig:overview_sne_PS20b}).

\subsubsection{1987A-like supernovae}
Only a small number of \Case B mergers end up producing stars with so massive envelopes to explode while appearing as a blue supergiant. The resulting 87A-like supernovae only account for $0.3\%$ of all core-collapse supernovae, with $M_\mathrm{ej}=11.5-33.1\Msun$. On average these transients can be quite energetic, as $E_\mathrm{kin,ej}=(1.0-5.1)\times 10^{51}\,\mathrm{erg}$. 

\subsubsection{Type IIn supernovae}

We predict that about $2.2\%$ of all core-collapse supernovae are \Type IIn. This number considers all models where \Case C RLOF is triggered  {after} core He depletion.  {We identify these progenitors in systems where either the donor star (either the primary or secondary star) triggers Case C RLOF with the companion, as well as merger products produced through inverse mass-transfer between the secondary star and an evolved primary that already depleted He in the core. This classification is made} regardless of how much material was lost during \Case C RLOF ($\Delta M_\mathrm{RLOF-C}$) to form the CSM. We can apply thresholds in $\Delta M_\mathrm{RLOF-C}$ above which to identify a model as a \Type IIn progenitor.  Within reasonable choices for the threshold of ($0.01\Msun$ or $0.1\Msun$) the number of identified \Type IIn supernovae would only slightly decrease (to $2.1\%$ and $1.9\%$ respectively). 

We distinguish the contribution to this population into three groups: those whose progenitors underwent stable \Case C RLOF ($26\%$ of all \Type IIn supernovae), unstable \Case C RLOF ($43\%$) and \Case B+C RLOF ($31\%$). Secondaries and merger products also contribute to the population, but by a negligible fraction (Fig.\,\ref{fig:overview_sne}) and we will neglect their properties in the following statistics.  The progenitors undergoing stable \Case C RLOF show larger $M_\mathrm{ej}=2.5-10.0\Msun$, and smaller $\Delta M_\mathrm{RLOF-C}\simle5.4\Msun$ compared to those undergoing unstable RLOF ($M_\mathrm{ej}=1.2-4.1\Msun$ and $\Delta M_\mathrm{RLOF-C}=6.6-7.0\Msun$,  {which also includes the mass of the whole H-rich envelope of the donor star before the onset of unstable mass-transfer as it also makes up the common-envelope}). The third group, made up by partially-stripped primaries following \Case B RLOF, shows lower-total masses by the time \Case C RLOF begins, and thus exhibit less mass-loss ($\Delta M_\mathrm{RLOF-C}\simle2\Msun$) while maintaining a wide range of ejecta masses ($M_\mathrm{ej}=1.4-8.3\Msun$). All \Type IIn supernovae show $E_\mathrm{kin,ej}=0.9-1.2\times 10^{51}\,\mathrm{erg}$, although this should be taken as an upper-limit since most of these progenitors come from low-mass models for which we do not have reliable estimates of the explosion properties (Sect.\,\ref{sec:methods:explodability}).

 \begin{figure}
    \centering
    \includegraphics[width=\linewidth]{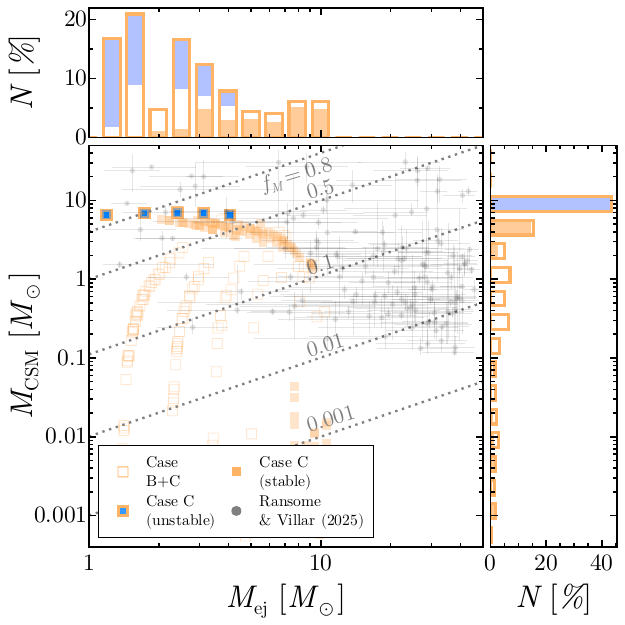}

    \caption{
     Ejecta masses $\Mej$ versus CSM masses $\Mcsm$ of our \Type IIn supernova progenitors (main panel) The right and upper panels show the histogram of $\Mcsm$ and $\Mej$ respectively. Lines decorating the scatter-plot indicate  constant $f_M$ (Eq.\,\ref{eq:fM}), which represents the conversion efficiency of the ejecta's kinetic energy into radiation as it impacts the CSM, assuming spherical symmetry. 
    We distinguish the contribution from models undergoing stable \Case C (orange filled), unstable \Case C (blue fill) and \Case C following \Case B (orange empty). The inferred properties from a sample of observed \Type IIn supernovae from \citet{RansomeVillar25_diversityIIn} is also shown (gray dots).}
    \label{fig:hist_IIi}
\end{figure}

\begin{figure}
    \centering
    \includegraphics[width=\linewidth]{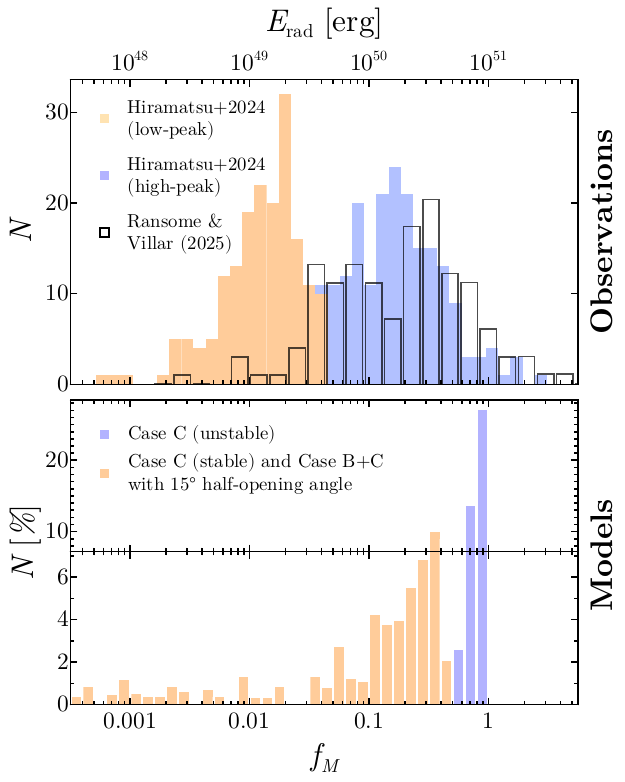}

    \caption{Comparison of the distribution of the integrated bolometric light-curve luminosity $E_\mathrm{rad}$ of observed \Type IIn supernovae (top) and the conversion efficiency $f_M$ from our models (bottom). The data from \citet{Hiramatsu25_IIn_bimodality} distinguish between a high-luminosity group ($E_\mathrm{rad}\sim1.8\times 10^{50}\,\mathrm{erg}$, blue) and a low-luminosity group ($E_\mathrm{rad}\sim1.4\times 10^{49}\,\mathrm{erg}$, orange), while that of \citet{RansomeVillar25_diversityIIn} is derived via \texttt{MOSFIT} by fitting and integrating the bolometric light-curve for each transient until day 200 after the explosion. Note that the data from \citet{RansomeVillar25_diversityIIn} is normalized to the total number of transients they analyzed. }
    \label{fig:hist_IIi_fm}
\end{figure}

We show the distribution of the \Type IIn supernovae in the $\Mej$-$\Mcsm$ plane assuming $\Mcsm=\Delta M_\mathrm{RLOF-C}$ in Fig.\,\ref{fig:hist_IIi}. The three groups we identified based on their evolutionary history occupy distinct regions in this diagram. Assuming that the CSM is distributed spherically around the progenitor, it follows that the group of unstable \Case C systems shows the strongest interaction (with conversion rates of $f_M=0.62-0.85$, see also the lower panel in Fig.\,\ref{fig:hist_IIi_fm}), followed by the other two groups which share similar $f_M<0.60$. 

The CSM may have a different geometry depending on whether mass-transfer was stable or unstable. The systems undergoing unstable RLOF may explode while the system is still in-spiralling (\PI) and therefore the CSM may appear as an extended, loosely bound and roughly spherical common-envelope. For the models undergoing stable RLOF, the unaccreted material may be ejected as outflows from the outer Lagrangian points (\citealt{Lu2023_L2_outflow_during_MT}, \PII) which may even remain bound to the inner binary \citep{ Pejcha16_CBD_LumRedNovae}. We assume that the models undergoing stable RLOF give rise to a flattened CSM, localized within an half-opening angle of $15^\circ$ from the orbital plane. 

Under this assumption, the distribution of $E_\mathrm{rad}$ is clearly bimodal (Fig.\,\ref{fig:hist_IIi_fm}, bottom panel; this bimodality is still visible even when using half-opening angles up to $30^\circ$ for stable mass-transferring systems). The high $E_\mathrm{rad}$ peak is made up of progenitors undergoing unstable \Case C RLOF (which we expect to develop a roughly spherical CSM) with, on average, $f_M(90^{\circ})\sim0.80$ and hence $E_\mathrm{rad}\simeq f_M(90^\circ)E_\mathrm{kin,ej}\sim8\times 10^{50}\,\mathrm{erg}$. The low $E_\mathrm{rad}$ peak, made up of systems undergoing stable \Case C or \Case B+C RLOF, is clearly distinguished from the first as $f_M(15^\circ)\sim0.1$ ($E_\mathrm{rad}\sim 10^{50}\,\mathrm{erg}$).

\subsubsection{Type Ibn supernovae}

Our fiducial model population yields that \Type Ibn supernovae make up $2.4\%$ of all core-collapse supernovae (Fig.\,\ref{fig:overview_sne}). As in the case of \Type IIn supernovae, this number includes all progenitors that undergo mass-transfer after core He depletion. If we include a threshold $\Delta M_\mathrm{RLOF-BB}\geq0.01\Msun$ to identify progenitors of \Type Ibn supernovae, the number of these transients would only diminish to $2.2\%$.  We note that the initial parameter space that produces \Type Ibn progenitors varies significantly for $M_\mathrm{1,i}=11.2\Msun, 12.6\Msun$ and $14\Msun$, which may be a sign that we could be under-resolving the progenitors' parameter space. We investigate this by interpolating all the models in the grid as a function of initial mass for fixed $P_\mathrm{i}$ and $\qi$ to identify missing progenitors of \Type Ibn supernovae for intermediate $M_\mathrm{1,i}$ not simulated in the grid, and found results that are in line with those without the interpolation.

\begin{figure}
    \centering
    \includegraphics[width=\linewidth]{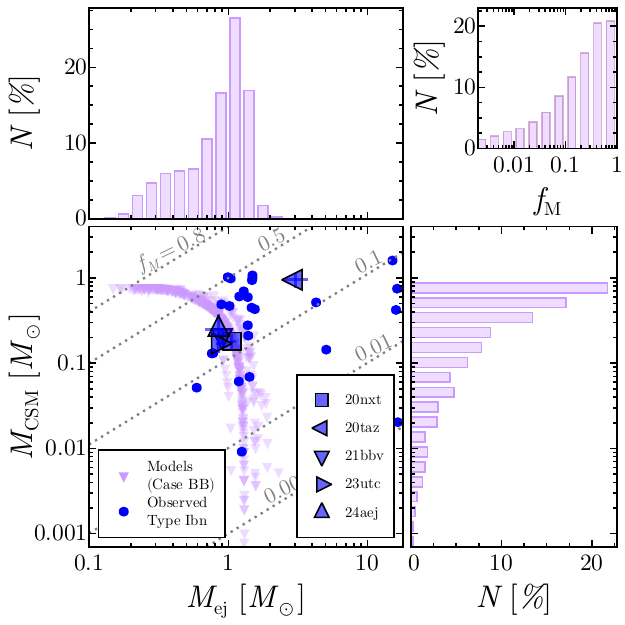}

    \caption{The same diagram as in Fig.\,\ref{fig:hist_IIi} for \Type Ibn progenitors. We include the inferred properties of the observed \Type Ibn  previously collected in \PII \ (see references therein, blue circles) including more recent ones from \cite{Wang25_5new_Ibn_SNe} (i.e., \object{SN2020nxt}, \object{2020tax}, \object{SN2021bbv}, \object{SN2023utc} and \object{SN2024aej}). On the top right, a histogram of $f_M$ is shown. }
    \label{fig:hist_Ibci}
\end{figure}
 
The progenitors of \Type Ibn supernovae have $\Delta M_\mathrm{RLOF-BB}\simle0.77\Msun$ (see Appendix\,\ref{appendix:CaseBB}) and $M_\mathrm{ej}=0.27-1.48\Msun$. While we also obtain $E_\mathrm{kin,ej}=10^{51}\,\mathrm{erg}$, we stress that it is likely an overestimate (see Sect.\,\ref{sec:methods:explodability} and \PII). Assuming $\Mcsm=\Delta M_\mathrm{RLOF-BB}$, we can construct a $\Mej-\Mcsm$ diagram in Fig.\,\ref{fig:hist_Ibci}. Assuming a spherically distributed CSM, we predict that our \Type Ibn progenitors would show $f_M\simle0.7$. 

Since these systems all underwent stable \Case BB RLOF, we can apply the same logic to that of stable \Case C and \Case B+C RLOF, whereby material is likely to be ejected close to the orbital plane, rather than being spherically distributed. Assuming that the material is found within an half-opening angle of $15^\circ$ from the orbital plane, we find $f_M(15^\circ)\simle0.2$ which would translate to $E_\mathrm{rad}\simle2\times 10^{50}\,\mathrm{erg}$, assuming explosion energies of $10^{51}\,\mathrm{erg}$. The values for $f_M$ and  $E_\mathrm{rad}$ are to be taken as upper-limit estimates, as we assume that the mass of the CSM is entirely made up from the mass-loss during \Case BB (while some part of the material may indeed escape the system and never interact with the supernova), and the explosion energy is likely overestimated (see above).

\subsubsection{Black holes}
We predict about $7$ BHs being formed for every 100 supernovae, with $M_\mathrm{BH}=7.5-19.9\Msun$ (i.e. the pre-collapse mass). This distribution shows a very long high-mass tail, extending up to $43\Msun$ which is produced by massive imploding mergers. The lower-mass ones ($\simle 10\Msun$) are for the most part produced by stripped primary stars.

\subsection{The effect of different explodability criteria}\label{sec:results:exp}

\begin{figure}
    \centering
    \includegraphics[width=1\linewidth]{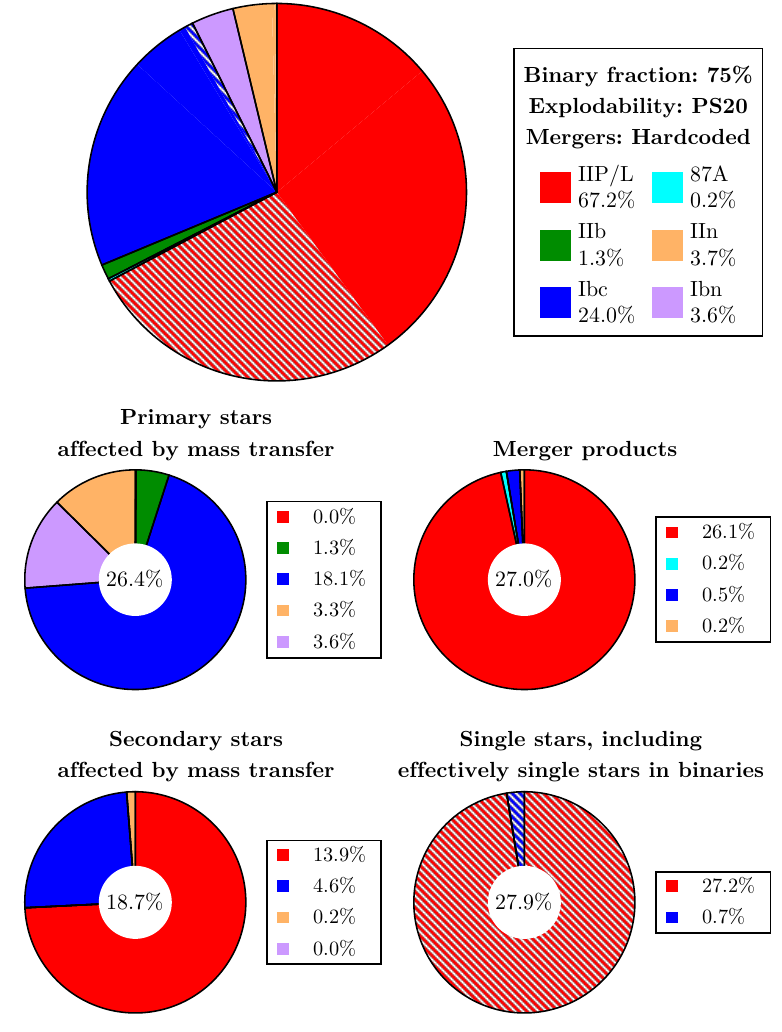}        
    \caption{As Fig.\,\ref{fig:overview_sne} but assuming the explodability criteria of \PS.}
    \label{fig:overview_sne_PS20a}
\end{figure}

\begin{figure}
    \centering
     \includegraphics[width=1\linewidth]{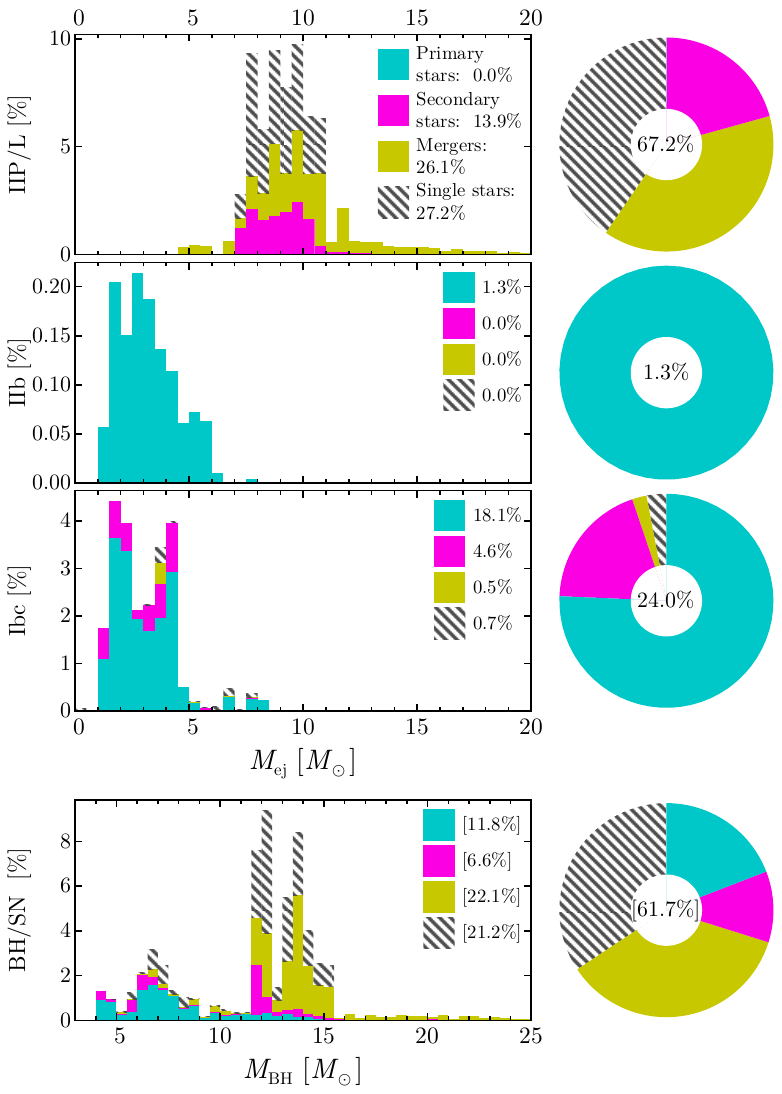}
    \caption{As Fig.\,\ref{fig:overview_sne_bin_typeb} but assuming the explodability criteria of \PS.}
    \label{fig:overview_sne_PS20b}
\end{figure}

\begin{figure*}
    \centering
    \includegraphics[width=.495\linewidth]{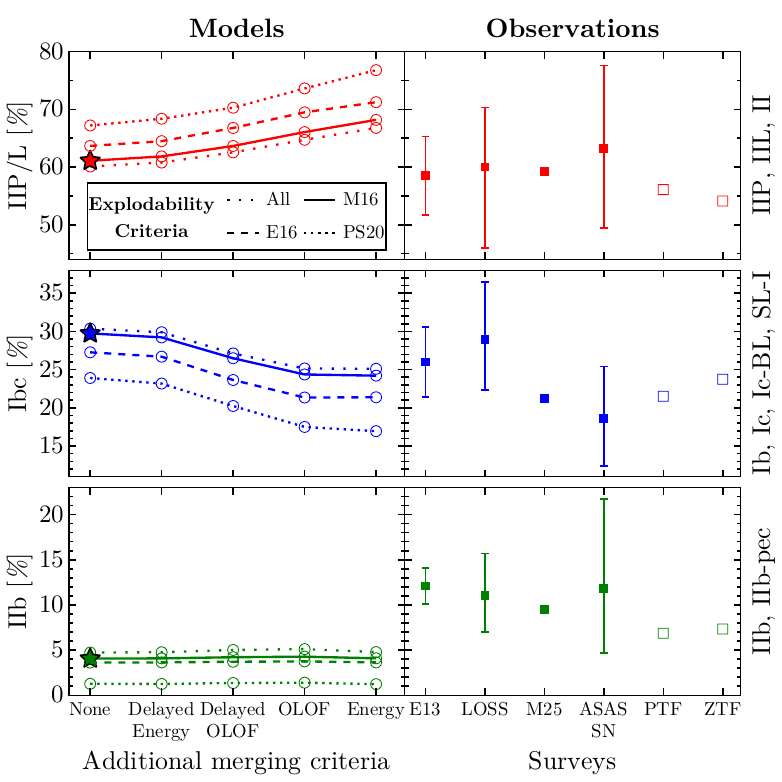}
    \includegraphics[width=.495\linewidth]{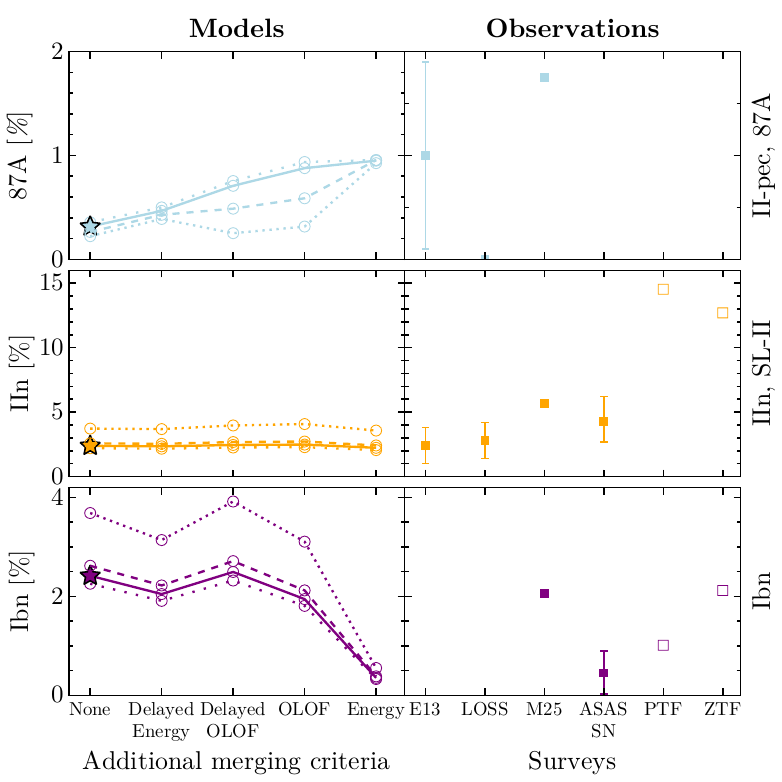}
    \caption{Comparison between the predictions on the fractions of supernovae of different types (different blocks) produced by the models (left sub-panel, with our supernova nomenclature on the left $y$-axis) and the observed fractions (right sub-panel, with the observed subtypes which we clump in each group on the right $y$-axis). The numbers are reported as a fraction of all core-collapse supernovae. 
    For the models, a line plot show the variation of the fraction of supernovae subtypes as a function of different merger criteria added on top of the ``Hardcoded'' criteria (`None' on the $x$-axis), sorted from left to right by those that produce less to more mergers that reach core-collapse (see text). Different dashing show the same while assuming different explodability criteria which are, sorted by increasing number of implosions, that all stars successfully explode (All), \citet[\Muller]{MHLC16}, \citet[\Ertl]{Ertl16_explodability_m4mu4} and \citet[\PS]{PS20_explodability_Mco_Xc}. The star corresponds to the fiducial model. For the observation, we show the estimates from volume-limited samples from \citealt{Eldridge13_deathmassivestarII_Ibc} (E13), the LOSS survey (specifically the subsample from high-mass galaxies, which serves as a proxy of high-metallicity environments,  \citealt{Graur2017_LOSS}), the sample from \citealt{Ma25_SNpops40Mpc} (M25) and from ASAS-SN \citep{Pessi25_ASASSN_volumelimitedrates} with filled-markers. We also include for reference magnitude-limited samples from PTF \citep{Schulze21_PTF_CCSNe} and ZTF \citep{ Hinds25_CSM_properties_II_ZTF}, shown with empty markers.}
    \label{fig:fractions_different_criteria}
\end{figure*}

Here we show the results from the fiducial model population (Sect.\,\ref{sec:results:example}) when adopting different explodability criteria than \Muller. For the effect on single-stars, refer to Appendix\,\ref{appendix:single_star}.

When applying the \PS \ method, the relative populations of \Type Ibc and \Type IIb supernovae are decreased. In Fig.\,\ref{fig:overview_sne_PS20a}, it is evident that these supernovae types lack contribution from the massive progenitors that were otherwise found in the fiducial population model (Fig.\,\ref{fig:overview_sne}). This is a direct consequence of the fact that in effectively single-stars, all progenitors with $M_\mathrm{pre-SN}\simgr12\Msun$ implode (Fig.\,\ref{fig:singleSN}) while most of the binary-stripped progenitors implode for $M_\mathrm{pre-SN}\simgr 10\Msun$ (Fig.\,\ref{fig:singleSN_HeS}). Now, the population of \Type IIb supernovae is exclusively made up of primary stars stripped in primaries, while also for \Type Ibc the contribution from self-stripped single stars is negligible. The ejecta masses are restricted to values below $\sim 8\mso$.

\Type IIP/L supernovae are instead mostly unaffected. This decreases the ratio of \Type Ibc to \Type IIP/L supernovae (Fig.\,\ref{fig:BHs_different_criteria}), but more so for the effectively-single star distribution than for the overall distribution (Fig.\,\ref{fig:overview_sne_PS20a}). Hence, the results over the whole distribution are more sensitive to different choices of $f_B$ as well as the upper limit on $P_\mathrm{i}$.

The number of interacting supernovae is indirectly affected, as they come from a limited mass-range where progenitors always explode, regardless of the explodability criteria adopted. In fact, the increasing number of imploding progenitors of other supernova types increases the relative number of \Type IIn and \Type Ibn supernovae by $\simgr1\%$ (Fig.\,\ref{fig:fractions_different_criteria}). Inevitably, this also increases the ratio of \Type Ibn to \Type Ibc by $\sim0.1$, while that of \Type IIn to \Type IIP/L is unchanged (Fig.\,\ref{fig:BHs_different_criteria}).

Because of the more imploding progenitors from both low and high masses, the number of BHs formed increases to $\sim60$ per 100 supernovae, and shifts the distribution of BH masses to $M_\mathrm{BH}=5.7-16.3\Msun$  (compared to $7.5-19.9\Msun$ when using \Muller).
Not only are BHs more numerous, but they also show a double-peaked distribution, with one peak around $4.4-8.6\Msun$ (about $24\%$ of the formed BHs) and the other around $11.6-18.1\Msun$ ($76\%$). Additionally, the high-mass tail in their distribution now extends to even higher masses, up to $55\Msun$.

With the explodability criterion from \Ertl, only the fractions of \Type IIP/L and \Type Ibc supernovae are significantly affected (Fig\,\ref{fig:fractions_different_criteria}, \ref{fig:BHs_different_criteria}) when compared to the fiducial model, but to a lesser extent than when employing the criterion of \PS. The number of BHs produced is $\sim16$ BHs per 100 supernovae,  with typical BH masses $M_\mathrm{BH}=7.0-15.3\Msun$.

\section{Comparison to observations}\label{sec:results:observations}

Several works have collected supernova observations and provided statistics on their distributions which can be compared to stellar models. Volume-limited samples are the prefered comparison tools, as they provide estimates that compensate for the magnitude bias. We compare our models with the samples from \citet[an updated sample from \citealt{Smartt2009_SNprogenitors_plusobs}]{Eldridge13_deathmassivestarII_Ibc}, the subsample from high-mass galaxies from the LOSS survey from \citet[initially reported by 
\citealt{Li2011_LOSS_SNe_sample} and reanalyzed in \citealt{Shivvers2017_LOSS}; the data in high-mass galaxies should correspond to high-metallicity environments, \citealt{Tremonti2004_MassMetallicity_relation_galaxies, Bevacqua24_galaxymass_metallicity_relation}, which better compares to our models]{Graur2017_LOSS}, the work by \cite{Ma25_SNpops40Mpc}, and the sample from ASAS-SN \citep{Pessi25_ASASSN_volumelimitedrates}.

\subsection{Type IIP/L, IIb and Ibc supernovae} 

The fraction of \Type IIP/L supernovae in our fiducial population model is within the ranges inferred by the observational samples (see Fig.\,\ref{fig:fractions_different_criteria}). The closest choice of model parameters that best falls in line with the average values reported in these studies are those with the least number of implosions (i.e. the criterion from \Muller) and the least number of mergers (i.e. the ``Hardcoded'' merger criteria). However, models with any combination of merger and explosion criteria adopted are still found within the error bars of the observations, with the exception of the model that combined the most mergers and the most implosions. \cite{Das25_LLIIP_CLU_ZTF} showed that the low-luminosity \Type IIP supernovae (referred to as LLIIP)
cannot be associated to single stars of mass between $8$ and $12\Msun$. In our models, thanks to binary-induced stripping, accretion and mergers, the population of \Type IIP/L supernovae originating from this mass range is reduced, which can ease this tension.

The observed fraction of \Type Ibc supernovae is less constrained, with some sources estimating around $\sim25-30\%$ (\citealt{Eldridge13_deathmassivestarII_Ibc},  \citealt{Graur2017_LOSS}) which yield good agreement with our models, preferentially those with less mergers and implosions (although the observational error-bars are still compatible with all models but those with the most mergers or most implosions). A similar conclusion cannot be drawn when comparing our models to the estimates from \cite{Ma25_SNpops40Mpc} and \cite{Pessi25_ASASSN_volumelimitedrates}, where the observations yield a small number of \Type Ibc supernovae ($\sim20\%$), which in this case falls better in line with models showing more mergers or more implosions. There is observational evidence for a few \Type Ibc supernova progenitors having been massive and likely not originating from binary-stripping (e.g., \citealt{Taddia19_broadlineIc,Karamehmetoglu23_highmass_Ibc}, but see \citealt{Dessart2020_Ibc} for possible degeneracies), which is compatible with our fiducial population model (see Sect.\,\ref{sec:results:example} and Fig.\,\ref{fig:overview_sne_bin_typeb}) but not with models in which all massive progenitors always implode (see Sect.\,\ref{sec:results:exp} and Fig.\ref{fig:overview_sne_PS20b}).

For the observed ratio of \Type Ibc to \Type IIP/L (Fig.\,\ref{fig:BHs_different_criteria}),  \cite{Eldridge13_deathmassivestarII_Ibc} and \cite{Graur2017_LOSS} estimate a value of $\sim0.45-0.48$, which is compatible with our models when producing less mergers and less implosions. The results of \cite{Ma25_SNpops40Mpc} and \cite{Pessi25_ASASSN_volumelimitedrates} yield a ratio of $0.30-0.35$, which instead agrees with models with more mergers or more implosions (Fig.\,\ref{fig:BHs_different_criteria}). 

\Type IIb supernovae make up about $10\%$ of the observed events in different volume-limited samples \citep{Eldridge13_deathmassivestarII_Ibc,  Graur2017_LOSS, Ma25_SNpops40Mpc, Pessi25_ASASSN_volumelimitedrates}, which is much greater than our predicted numbers regardless of the chosen set of parameters (Fig.\,\ref{fig:fractions_different_criteria}). This could be connected to the winds mass-loss recipes adopted in the stellar evolutionary models which strip low-mass H-rich envelopes too efficiently (see Appendix\,\ref{sec:discussion:winds_and_Z}) or inconsistencies in the classification of \Type IIb supernovae in models and observations. Moreover, the bulk of \Type IIb supernovae is expected to come from stars that underwent binary interaction \citep{Smith2011_obsSNfractions} while our fiducial population models suggest the opposite. There are nonetheless observations of \Type IIb supernovae suggestive of a high-mass progenitor \citep[e.g.,][]{RubinGalYam16_TypeIISNe_LCs}.

Our estimates for the fraction of 87A-like supernovae are compatible with those from \cite{Eldridge13_deathmassivestarII_Ibc} and the non-detections from high-mass galaxies from \cite{Graur2017_LOSS}. \cite{Ma25_SNpops40Mpc} instead found a fraction of 87A-like supernovae closer to $2\%$, which is distinctively larger than even our more optimistic estimate when assuming more mergers. While this fraction may be larger if one includes secondary and primary stars (which we excluded because of MLT++, see Sects.\,\ref{sec:methods:sn_type} and Appendix\,\ref{sec:discussion:singlestar}), this would additionally decrease the number of \Type IIb supernovae, which are already in tension with observations.

\subsection{Type IIn supernovae}

The works from \cite{Eldridge13_deathmassivestarII_Ibc}, \cite{Graur2017_LOSS} and \citep{Pessi25_ASASSN_volumelimitedrates} indicate the relative number of \Type IIn to be in the order of $2-4\%$ of all core-collapse supernovae, which is compatible with our estimates. 
The more recent work from \citealt{Ma25_SNpops40Mpc} instead reports a number of \Type IIn supernovae closer to $5.6\%$ of all core-collapse supernovae, about a factor two higher than our predictions.    

A large-scale attempt to fit numerous \Type IIn supernova light-curves was made in \cite{RansomeVillar25_diversityIIn} by analyzing 147 transients using the code \texttt{MOSFIT} \citep{Chatzopoulos12_CSM_SLSN, Mosfit_2018}. Their analysis revealed that most of the observed \Type IIn supernovae have $\Mcsm=1.26^{+6.68}_{-0.92}\Msun$ and $\Mej=20.1^{+19.0}_{-14.9}\Msun$. These values are in tension with those derived within the framework of our models (see Fig.\,\ref{fig:hist_IIi}), although there are events that singularly overlap with our models. 
Such massive events may be the result of a non-volume-limited sampling, which may be biased towards more luminous and extreme events  {likely coming from more uncommon evolutionary channels like pulsational pair-instability \citep[PPI,][]{Woosley2017_PPISN}, or LBV-like eruptions \citep[e.g., for the progenitor of \object{SN2010jl},][]{Niu24_2010jl_LBV} which we do not consider in our models}.

We caution that the assumption of spherical geometry for modeling CSM-interaction in \texttt{MOSFIT} may affect the inferred fitting parameters for interacting supernovae where the CSM is asymmetrically distributed. Several observations of \Type IIn supernovae such as \object{SN2012ab} \citep{Bilinski18_2012ab_IIn_asphericCSM}, \object{SN2013L} \citep{Andrews17_2013L_IIn_asymmCSM} and \object{SN2020ywx} \citep{BaerWay25_2020ywx_IIn_asymmCSM} highlight an asymmetric CSM at the time of explosion. This has also been observed for the \Type II \object{SN2023ixf} \citep{Vasylyev25_2023ixf_polarization}. Surveys of \Type IIn supernovae via spectropolarimetry \citep{Bilinski24_IIn_spectropol_asymmetricCSM} and UV and Optical studies of their light curves \citep{Soumagnac20_UV_IIn_asphericCSM} suggest that strong asymmetry in the CSM is found for $\sim35-66\%$ of all \Type IIn supernovae. This hints that perhaps all \Type IIn supernovae may have asymmetric CSM when considering the random viewing angle distribution of the observer with respect to the CSM \citep{Bilinski24_IIn_spectropol_asymmetricCSM}.
Other events, such as \object{SN2014C}, can be explained in the context of interaction with a toroidal CSM \citep{Bietenholz21_asymmetry_2014C}  {which can be produced in a thousand year long enhanced mass-loss through a binary companion \citep{Milisavljevic15_SN2014C, Orlando24_2014C_CSM_Torus} without necessarily being driven by common-envelope evolution, as stable mass transfer can still provide a bound, massive, and aspherical CSM.} 

A more comprehensive sample of 475 \Type IIn supernovae has been reported in \citet{Hiramatsu25_IIn_bimodality}. They note that the transients show a bimodal distribution of their integrated light-curve luminosities, with one group showing $E_\mathrm{rad}\sim 1.4\times 10^{49}\,\mathrm{erg}$ and a second group with $E_\mathrm{rad}\sim 1.8\times 10^{50}\,\mathrm{erg}$. This bimodality is qualitatively similar to that we find in the models (see Fig.\,\ref{fig:hist_IIi_fm}), although our peaks are found at higher $E_\mathrm{rad}$ (assuming $E_\mathrm{kin,ej}\sim 10^{51}\,\mathrm{erg}$ in our models). This difference is not surprising, as we assume that all of $\Delta M_\mathrm{RLOF-C}$ interacts with the supernova. As some of it may be unbound from the binary and never interact with the supernova, the effective value of $f_M$ may be substantially lower and hence offer a way to reconcile our $E_\mathrm{rad}$ with the observation. We also note that the relative number of systems undergoing unstable versus  {stable \Case C plus \Case B+C RLOF} (43\% to 57\%) is not dissimilar from that of low- versus high- luminosity groups in \cite{Hiramatsu25_IIn_bimodality} (51\% to 49\%), even though the observations do not provide a volume-limited sample and hence may over-represent high-luminosity transients.

Finally, our models do not predict the occurrence of outburst prior to the explosion, as observed in many \Type IIn supernovae \citep{Ofek14_preSN_outbursts_IIn, Strotjohann2021_interactionpoweredSN, Reguitti24_precursors_of_IIn}. We refer to Appendix\,\ref{sec:discussion:outbursts} for a discussion. 

\subsection{Type Ibn supernovae} 

\cite{Hosseinzadeh2017_Ibn_lightcurve} collected \Type Ibn supernovae to construct a template light-curve of these objects, with which one can derive typical radiated energies of these transients to be $E_\mathrm{rad}=(1.6-3.8)\times 10^{49}\,\mathrm{erg}$. This value is about a factor three smaller than what we predicted from the models when assuming the CSM focused closer to the orbital plane. However, significantly lower explosion energies (\PII) and lower CSM masses in the models can push $E_\mathrm{rad}$ closer to the observed values. 

For some of the observed supernovae, light-curve fitting has produced estimates for CSM and ejecta masses.  {Much like the case of \Type IIn supernovae, there are some outliers with exceptionally high energies and/or high inferred ejecta masses, which could be produced through PPI \citep{Woosley2017_PPISN}. Besides this,} the bulk of these observations seem to coincide with the parameter space predicted by our models (see \PII). Interestingly, the region occupied by most observations in Fig.\,\ref{fig:hist_Ibci} coincides with the more populated one theoretically (within the lines $f_M=0.1$ and $f_M=0.5$). Despite this, we nonetheless remain cautious of the inferred masses from these events, since some of them show signs of asymmetry in the CSM \citep[e.g., \object{SN2023fyq},][]{Dong2024_SN2023fyq_Ibn_eqdisk}. The asymmetry would likely affect the estimates of $\Mcsm$ and $\Mej$ from light-curve fitting, since spherical symmetry is assumed in these models.

Until recently, estimates on the fraction of \Type Ibn supernovae with respect to the population of core-collapse supernovae were not available. Works from PTF \citep{Schulze21_PTF_CCSNe} and ZTF \citep{Perley2020_ZTF_SNdemographics, Hinds25_CSM_properties_II_ZTF}, report that \Type Ibn supernovae seem to contribute about $1-3\%$ of all supernovae within a magnitude-limited sample. The first volume-limited estimates of \Type Ibn supernovae come from \citet{Ma25_SNpops40Mpc} and \cite{Pessi25_ASASSN_volumelimitedrates}. \cite{Ma25_SNpops40Mpc} suggests that \Type Ibn supernovae contribute to about $2\%$ to all core-collapse supernovae  (Fig.\,\ref{fig:fractions_different_criteria}), with a ratio of \Type Ibn to \Type Ibc of about $0.1$. While the uncertanties on this number are quite large, it is compatible with the results of our fiducial model. Both the number of \Type Ibn supernovae and their ratio to \Type Ibc is incompatible with the population models assuming the ``Energy'' criterion for mergers or the \PS \ criterion for explodability. Besides this, the other merger and explodability criteria combinations produce fractions of \Type Ibn supernovae compatible with the observed value.

\cite{Pessi25_ASASSN_volumelimitedrates} instead observe only $0.5\%\pm0.4\%$ of all core-collapse supernovae as \Type Ibn. Such small fraction is only found in models with the ``Energy'' merger criterion, in direct contrast to the conclusions inferred from comparing models to the results of \cite{Ma25_SNpops40Mpc}.

\section{Conclusions}\label{sec:conclusions}

In this work, we have found that about two-thirds of all \Type IIP/L supernova progenitors are expected to have experienced mass accretion or merging. Therefore, binary evolution can lead to much higher envelope masses and ejecta masses, compared to single stars (Fig.\,\ref{fig:overview_sne}). Our models also predict a small population of \object{SN1987A}-like transients from massive merger products based on a high final envelope-to-core mass ratio. 

Binary evolution naturally enhances the number of stripped-envelope supernovae (\Type Ibc and \Type IIb),
and leads to a number ratio of Type\,Ibc to Type\,IIP/L which is in rough agreement with the observed one. In particuar, it enables a copious amount of low-mass \Type Ibc and \Type IIb supernovae, via binary-stripped primaries and secondaries (Fig.\,\ref{fig:overview_sne_bin_typeb}). 

Our models also show a strong dependence of the progenitor and ejecta mass distribution of stripped-envelope supernovae from the employed explodability criteria. We find that using the \Muller \ criteria leads to supernova population which is very similar to that found when assuming that all massive stars explode (as it only forms seven BHs per 100 supernovae). This leads to most Type\,IIb and one-third of \Type Ibc supernovae showing ejecta masses of
$\simgr 12\mso$ (Fig.\,\ref{fig:overview_sne_bin_typeb}). With the explodability criteria of \PS, on the other hand, we obtain 60 BH per 100 SNe, and the massive progenitors of \Type IIb and \Type Ibc all implode (Fig.\,\ref{fig:overview_sne_PS20b})

Our comprehensive and dense grid of binary evolution models allows also for the first assessment of the expected population of interacting supernovae, in the frame of our assumption that a core-collapse supernova arising during mass transfer gives rise to this phenomenon (\PI, \PII). 
Our models predict that \Type IIn supernovae make up  $2.4\%$ of all core-collapse supernovae, consistent with volume-limited supernova samples \citep{Eldridge13_deathmassivestarII_Ibc, Graur2017_LOSS}. Similarly, \Type Ibn supernovae make up $2.4\%$ of all core-collapse supernovae, which is compatible to the value derived in the volume-limited sample in \cite{Ma25_SNpops40Mpc}.
Furthermore, our models show a bimodal distribution of radiated energies of Type IIn events, which qualitatively agrees with observations \citep{Hiramatsu25_IIn_bimodality}, when assuming that stable mass transfer produces disk-like CSM structures while unstable mass transfer yields more spherical outflows.

The high-cadence observations of transient from the concurrent operations of LSST and ZTF, along with continued analysis of archival data from current programs, will provide invaluable insights into the populations of observed supernovae.  In particular, focused efforts on identifying and characterizing interacting supernovae will provide critical constraints on the final phases of stellar evolution and the nature of mass loss near core-collapse.

\begin{acknowledgements}
This project made use of the Julia language \citep{code_Julia2017}. All plots were made with the Makie.jl package \citep{code_Makie2021}. The authors thank the anonymous reviewer for their valuable and constructive feedback, which has significantly contributed to the enhancement of this manuscript. The authors additionally express their gratitude to Luc Dessart for inspiring ideas and comments, as well as to Adam Burrows and Qiliang Fang for constructive comments. AE thanks Fabian Schneider for sharing data from his models and for helpful discussions. AE is also grateful for many insightful discussions with Eva Laplace, Philip Podsiadlowski, Vincent Bronner,  Ruggero Valli, Alexander Heger, Ylva G\"otberg, Reinhold Willcox, and Bethany Ludwig.

\end{acknowledgements}

\bibliographystyle{aa}
\bibliography{Reference_list}

\begin{appendix} 

\section{Model uncertainties}\label{sec:discussion}

In the following, we outline and discuss the shortcomings of our models and compare our results with previous works. We will not cover in detail the uncertainties linked with stellar and binary physics (see instead \citealt{Jin2024_boron}, \citealt{Jin25_BonnGal}, as well as \PI \ and \PII \ for more details) but rather discuss only their effect on the population estimates of the different transients.

\subsection{Stellar and binary physics}\label{sec:discussion:model_physics}
\subsubsection{Winds and metallicity}\label{sec:discussion:winds_and_Z}
Stellar winds are an important factor in determining if single and partially-stripped stars retain a massive H-rich envelope at the time of core-collapse. This influences both the ratio of \Type IIP/L to \Type Ibc supernovae as well as the absolute number of \Type IIb supernovae. Different metallicities also affect the winds, as they typically scale with metallicity (\citealt{Mokiem2007_Z_winds_OBstars, Hainich15_winds_Z_WRs_SMC, Backs24_winds_Z_XSU}, except perhaps for the winds of red supergiants, see e.g. \citealt{Kee_Atmo_Pturb_2021,Antoniadis24_RSGmasslossZ, Schootemeijer24_no_companions_for_WRs_SMC}).

Since in our models all the progenitor channels (i.e. primary stars, secondaries, merger products and single stars) produce a similar proportion of \Type IIb SNe ( $4-5\%$, see Fig.\,\ref{fig:overview_sne}), which is a factor two smaller than what inferred in observations (Sect.\,\ref{sec:results:observations}), this implies that we adopt wind mass-loss recipes that strip low-mass H-rich envelopes too efficiently. Recent works indeed suggest that mass-loss rates for partially-stripped stars may actually be lower than the rates we adopt \citep{Sravan_b, Gilkis_wind, Gilkis22_H_in_IIb_Ib, Vink_WR, Gotberg2023_stripped_stars_in_binaries}. 

Adopting lower mass-loss rates for partially-stripped stars would shift many of our low-mass \Type Ibc progenitors to \Type IIb. Its effect on \Type Ibn progenitors is mixed (see \PII), as it would shift the parameter space of the progenitors to lower-masses (which would be favored by the IMF) but it would also likely convert many of these to \Type IIb or even \Type IIn. Finally, it may increase the number of \Case B+C systems (and hence \Type IIn supernovae) due to the donor stars retaining more massive envelopes following \Case B, which can expand significantly enough to drive \Case C RLOF later on.

 Lower metallicities would reduce opacities in addition to the wind mass-loss rates, and therefore shrink the parameter space for RLOF. This would on the one hand decrease the number of supernova progenitors that underwent RLOF, which would decrease the number of \Type Ibc and the interacting supernovae. Combined with the lower-winds, it likely increases the population of \Type IIb supernovae  (see \citealt{Souropanis25_SESNe_Metallicity_POSYDON}), while its effect on \Type IIn and \Type Ibn supernovae is non-trivial .

\subsubsection{The treatment of convection}\label{sec:discussion:singlestar}
The results shown in this work are sensitive to the evolution of the donor star's radius, which is affected by the assumptions on the mixing-length parameter $\alpha_\mathrm{MLT}$ (which we fix to $1.5$) as well as the treatment of sub-surface convection \citep[see ][]{MESA_III}. Different choices will lead to differences in the onset and the evolution of mass-transfer, therefore impacting the parameter space for mass-transferring binaries as well as the appearance of interacting supernovae.

Previous works \citep{Dessart13_IIPSNe,Goldberg19_IIPSNe, Dessart2024_partiallystripped_typeII_progs} suggest that the choice of $\alpha_\mathrm{MLT}$ should be higher than what we prescribe in our models (and hence reduce the radius of the progenitors), due to the lack of observed \Type IIP supernovae with large progenitor radii (in most observed \Type IIP supernovae, \citealt{Martinez22_CSP_hydromodelling_obs_IIP} suggest progenitor radii of $R\simle 800\Rsun$, while our models predict radii of up to $1\,500\Rsun$). Were we to adopt a higher $\alpha_\mathrm{MLT}$, this would reduce the number binary systems undergoing mass transfer, which would affect the supernova population by diminishing \Type Ibc and interacting supernovae.

One of the main side-effects of the use of MLT++ (Sect.\,\ref{sec:method:stellar_physics}), is to shrink the radii of stars with high $L/M$. This can be observed by direct comparison of the models with $M_\mathrm{1,i}=12.6\Msun$ in the Bonn-GAL grid (Fig.\,\ref{fig:logPq_1.10}) and those shown in \PI, where all physics are the same except MLT++ which was not implemented. We note that the parameter space of \Type IIn supernovae is larger in \PI \ than those in the Bonn-GAL grid with $M_\mathrm{1,i}=12.6\Msun$ by $22\%$.  
Disregarding MLT++ would also lead to slightly more mass-loss during \Case B RLOF, resulting in less \Type IIb supernova progenitors being produced (as they would also undergo \Case C RLOF) which would increase the already existing tension with the observations (Sect.\,\ref{sec:results:observations}). 

\subsubsection{Extended atmospheres, pulsations and orbital eccentricity}
There are other physical processes that are neglected in the stellar evolution models that will affect the observed number of interacting supernovae. In \PI \ and \PII, we identified overall three processes that would affect \Type IIn and \Type Ibn supernovae.

For wide binaries, the red supergiant may show a quasi-hydrostatic extended atmosphere \citep[as observed in galactic red supergiants][]{ArroyoTorres_whylargeRSG} which may help trigger \Case C RLOF in much wider binaries than found in the stellar evolutionary models used here (\PI) and thus increase the number of \Type IIn supernovae. A pulsationally-unstable progenitor \citep{Sengupta25_envelope_pulsations_eruptions_RSGs, Ma2025_AREPORSG} or an eccentric binary may also drive additional mass-loss through periodic mass-transfer.  The sum of these effects could help reconcile the differences between our estimated fraction of \Type IIn supernovae and those reported in \cite{Ma25_SNpops40Mpc} (Sect.\,\ref{sec:results:observations}). It is overall unclear the direction in which all other supernova progenitors, especially those of \Type Ibn supernovae, would be affected.

\subsubsection{Mass transfer}

The amount of mass accreted by the companion in our models is limited by the rotation rate of the accretor. In all systems except the tighter and more massive \Case A, the secondary rarely accretes more than $0.1-0.3\Msun$ of material. The real accretion efficiency may be higher (see \citealt{Vinciguerra2020,Schootemeijer18_PhiPersei, Xu25_SMC_pop_mesa, Schuermann25_SMC_pop_combine} as well as the discussions in \PI \ and \PII).  

Higher accretion efficiencies would produce more massive secondaries, more massive mergers (as less mass was lost from the system prior to the onset of unstable mass-transfer) and slightly less massive donors \citep{Claeys_b}. This would increase the number of \Type IIP/L supernovae significantly, as low-mass secondaries may accrete enough to become \Type IIP/L supernovae (more so than \Type Ibc or even BHs due to the IMF). At the same time, mass-transfer may be more prone to turn unstable due to the response of the accretor \citep{Braun_Langer_95, SchuermannLanger25_stableRLOF_accretor} and thus lead to less stripped- or partially-stripped supernovae (which would have been produced by primaries). 
Rapid population synthesis works from \cite{Kinugawa24_popsynth3_SNe_binaries} and \cite{Ko25_PopSynth_Ibn} find that their numbers of \Type IIP/L, \Type Ibc and \Type Ibn supernovae do not significantly change with different accretion efficiencies, but are instead more sensitive to the assumptions on the outcome  of the common-envelope phase.

\subsection{Population synthesis assumptions}\label{sec:discussion:postprocess}
\subsubsection{The effect of different merger criteria}\label{sec:results:parameters}

The different merging criteria change the number of models labeled as mergers. About $32.2\%$ of core-collapsing stars are merger products when adopting the  ``Hardcoded'' criteria. 
The other criteria (see Sect.\,\ref{sec:methods:binary_physics}) produce increasing fractions of core-collapsing stars coming from merger products, namely the ``Delayed Energy'' produces $33.8\%$ of them, the ``Delayed OLOF'' criterion $38.1\%$, the ``OLOF'' criterion $41.5\%$ and finally the ``Energy'' criterion $44.1\%$. 
The number of core-collapse supernovae slightly increases with more stringent merger criteria (as more low-mass binaries where both components could not reach core-collapse can now merge into a more massive star that can), by at most by $5\%$ when compared to the ``Hardcoded'' criteria.

With more stringent merging criteria, the fraction of \Type IIP/L supernovae increases as merger models preferentially explode as \Type IIP/L supernovae while one of the two stars was likely a \Type Ibc progenitor if it had not merged (Fig.\,\ref{fig:overview_sne}). Naturally, the number of \Type Ibc supernovae decreases, which also implies a decrease in the the ratio of \Type Ibc to \Type IIP/L supernovae (by as much as $-0.15$, see Fig.\,\ref{fig:BHs_different_criteria}). 

Increasing the number of mergers does not significantly increase the number of 87A-like supernovae, which reach at most $1\%$ with the  ``Energy'' criterion. These supernovae require over-massive H-rich envelopes that are only produced in the merger of a post main-sequence primary with a massive H-rich companion during \case B RLOF. In our grid, such systems are either stable across all merger criteria or undergo inverse-mass transfer after the primary has developed into a He star. 

Implementing various merging criteria does not alter the fraction of \Type IIb supernovae (Fig.\,\ref{fig:fractions_different_criteria}), as the initial parameter space of their progenitors is not flagged by any of these criteria (e.g., Fig.\,\ref{fig:logPq_1.10}). A similar statement can be made for \Type IIn supernovae, although we still count systems undergoing unstable mass transfer as producing interacting supernova progenitors (see Sect.\,\ref{sec:method:interaction}). The number of BHs produced and their masses are not significantly affected by changing merging criteria.

We find that the fraction of \Type Ibn supernovae changes dramatically only when using the ``Energy'' criterion, where the fraction is reduced to only $\sim0.5\%$ of all core-collapse supernovae. This follows because these progenitors originate in systems with lower $M_\mathrm{1,i}$, $\qi$, and $P_\mathrm{i}$, which are generally more prone to be flagged as mergers with the ``Energy'' criterion.

\subsubsection{Varying distributions}\label{sec:discussion:distributions}
Our analysis exclusively assumed a binary fraction $f_B=75\%$ (with all orbits being circular) and used a \citet{Salpeter_IMF_55} IMF ($\alpha=-2.35$), a flat $\log P$ and $q$ ($\beta=\gamma=0$) distributions (see Eq.\,\ref{eq:distribution}). Varying these parameters offer quantitative differences that do not affect the main message of this work. We present the results by varying these distributions, assuming the same explodability and merger criteria of the fiducial model (Sect.\,\ref{sec:results:example}), where the results will be presented as differences from those in the fiducial model.

A top-heavy IMF (assuming $\alpha=-1.85$), mainly increases the ratio of \Type Ibc to \Type IIP/L supernovae ($+0.24$), as more massive stars are being produced which typically explode as \Type Ibc rather than \Type IIP/L, and decreases the number of Interacting supernovae ($-1.1\%$), as they are produced in systems with low initial mass. A bottom-heavy IMF ($\alpha=-2.85)$ has the mirrored effect with shifts of a similar magnitude. 

With $\beta<0$, we increase the contribution from systems with small $P_\mathrm{i}$ and decrease that from systems with higher $P_\mathrm{i}$. Assuming  $\beta=-0.75$ \citep{Sana_massive_stars_binaries}, 
the number of merger progenitors increases by $+18.5\%$ as the lowest-period binaries always end up merging. As a consequence, the number of \Type IIP/L supernovae increases while \Type Ibc decreases, resulting in a lower \Type Ibc to \Type IIP/L ratio ($-0.09$). \Type IIn progenitors drop by a factor two, as they originate in wide-binaries ($-1.3\%$) while \Type Ibn and \Type IIb are mostly unaffected.

For  $\gamma=-0.65$ the contribution from mergers increases ($+7.1\%$), as low-mass ratio systems always merge, resulting in more \Type IIP/L supernovae and less \Type Ibc, hence yielding a smaller ratio of \Type Ibc to \Type IIP/L by $-0.10$. Similarly,  \Type IIb, \Type IIn and \Type Ibn each slightly diminish ($\simle-0.5\%$).  For $\gamma=+0.48$, the opposite occurs. 

There is evidence that high-mass stars show a higher binary fraction of $\simgr90\%$ above $M_\mathrm{1}\sim10\Msun$ \citep{Sana14_Ostar_binaries,MoeDiStefano2017,Offner2023_multiplicity}. We therefore studied the population of supernovae assuming $f_B=100\%$, and find that overall \Type IIP/L supernovae diminish by $\sim5\%$ and \Type Ibc increase by a similar margin, which increases the \Type Ibc to \Type IIP/L ratio by $0.11$. \Type IIb supernovae are mostly unaffected, while interacting supernovae increase (with both \Type IIn and \Type Ibn each increasing by $+0.6$). It is important to note that even with $f_B=100\%$, about $13.5\%$ of all core-collapse supernovae originate from effectively single stars, of which $58\%$ of are \Type IIP/L, $39\%$ \Type Ibc and $3\%$ \Type IIb. 

\subsubsection{Merger products and evolution after core-collapse}\label{sec:discussion:postSN_mergers}
The evolution of merger products and secondaries relied on simplifying assumptions (see Sects.\,\ref{sec:methods:megers},\,\ref{sec:method:postSN}). Since these progenitors make up a significant fraction of core-collapse supernovae ($29.6\%$ and $16.4\%$, Sect.\,\ref{sec:results:example}) the use of different assumptions will affect the results shown in this paper.

 Since we assume that all models undergoing unstable mass-transfer merge, the fraction of \Type IIP/L supernovae is likely an upper-limit. If many were to eject the CE instead of merging, they would form a tight He-star + main-sequence binary which could result in a \Type Ibc or even \Type Ibn supernova, depending on the orbital evolution after the common-envelope is ejected. For stars that do merge, previous works also argue that part of the H-free core may be dug up, \citep{MenonHeger17_mergers, Menon24_mergermodels} which may help convert \Type IIP/L progenitors into 87A-like progenitors. 

The evolution of secondary stars is heavily influenced on whether or not it remains in a close binary with the remnant following the explosion of the primary. Previous works show that around  $\sim20\%$ of binaries are expected to remain bound after a successful explosion \citep{ DeDonder97_Runaways_kicks, BrandPodsi95_SNkicks_and_NSbinaries, Eldrige11_runaway_progenitors_SNe, Renzo19_runaway_postSNbreakup}, which would preferentially affect systems where the companion is more massive than the exploding star and where the orbit is tighter. This translates to models with high $\qi$ and low $P_\mathrm{i}$ where the primary is stripped through either \case A or \Case B RLOF. In the context of our models, secondary stars in these systems would typically produce \Type IIP/L supernovae. If such progenitors remain in a tight-enough orbit with a NS to trigger RLOF in later evolutionary phases, it will likely develop into a T$\dot Z$O and be disrupted, thus also decreasing the number of \Type IIP/L supernovae. 

Our simplifying assumptions for the stability of mass-transfer between the secondary star and the BH remnant will also affect the population of supernovae. When comparing our BH + main-sequence star binaries with the models in \cite{Klencki25_OB_BH_binaries_and_BBH_mergers}, (more precisely those with $6\Msun\simle M_\mathrm{BH}\simle13\Msun$, which are typical masses in binaries where the secondary will transfer mass to the BH), we note that our assumption in Sect.\,\ref{sec:method:postSN} is roughly  recovered. In our fiducial model, we mistakenly label $\simle 6\%$ of mass-transferring models with a BH as undergoing stable/unstable mass-transfer compared to the models in \cite{Klencki25_OB_BH_binaries_and_BBH_mergers}, which does not affect our statistics significantly. When adopting the explodability criterion from \Ertl \ or \PS, the absolute number of systems undergoing a different evolution from those in \cite{Klencki25_OB_BH_binaries_and_BBH_mergers} increases two- and four-fold respectively, which is still not enough not to affect the population of supernovae significantly.

\subsection{Pre-supernova outbursts}\label{sec:discussion:outbursts}

Our models offer one way to produce CSM in late evolutionary phases, which may be the (or one of the) main channel(s) to produce interacting supernovae. While we can replicate the number of \Type Ibn and \Type IIn supernovae under certain assumptions on the explodability and merger criteria, these still fall short of explaining some of the observed features of some transients.

There are observations of outbursts preceding the explosion of both \Type Ibn and \Type IIn supernovae by a few months or years \citep[e.g.,][]{Strotjohann2021_interactionpoweredSN}.  Several works have shown that a significant fraction of the progenitors of \Type IIn supernovae show pre-explosion outbursts \citep[$\simgr 10-50\%$, ][]{Ofek14_preSN_outbursts_IIn, Strotjohann2021_interactionpoweredSN, Reguitti24_precursors_of_IIn}, with some showing features compatible with those observed during the eruptions of LBVs \citep{Reguitti24_precursors_of_IIn}.  It is less clear to what degree these outbursts are occurring within \Type Ibn supernovae, as some transients clearly lack bright outbursts (e.g., \object{SN2015G}, \citealt{Shivvers17_2015G_Ibn}, or \object{SN2020nxt}, \citealt{Wang24_2020nxt_Ibn_noprecursor}). However, the observation of sustained asymmetry in some interacting supernovae may be harder to reconcile with the occurrence of pre-supernova outbursts and is better explained with sustained mass-loss through binary interaction \citep{Bilinski24_IIn_spectropol_asymmetricCSM}. Similarly, these outbursts alone may not be able to explain the observed narrow-line features for all observed events \citep{Strotjohann2021_interactionpoweredSN}.

Some pre-supernova activity may be expected by our \Type IIn progenitors that undergo unstable \Case C RLOF. Here, as the system develops a common-envelope, the short-timescale orbital decay that follows injects a significant energy to the common-envelope, and may produce a Luminous red nova \citep{SokerTylenda2003_V838Mon_LRN, Ivanova13_LRT_CEE, Pastorello19_LRN_mergers_or_eruptions, HatfullIvanova25_simulatingCEE_LRN} which may be triggered in our models up to a few thousand years before the supernova explosion.

In all other instances, our models do not predict outbursts. It is likely however that binary-induced mass-loss may help trigger pulsations and variability as the star is pushed closer to the Eddington limit \citep[as it increases the $L/M$,][]{Heger97_RSG_LM_pulsations_CSM}, which is thought to be the case for \object{SN2012ab} \citep{Bilinski18_2012ab_IIn_asphericCSM}. For \Type Ibn supernovae, a late unstable phase of mass-transfer may occur in the last months before core-collapse (\Case X RLOF, see \citealt{WuFuller_IbcLMT}, \PII) which could appear as an outburst. However, the fraction of supernovae affected by this is unclear.

\FloatBarrier

\section{Single star evolution}\label{appendix:single_star}

\begin{figure}
    \centering
    \includegraphics[width=0.99\linewidth]{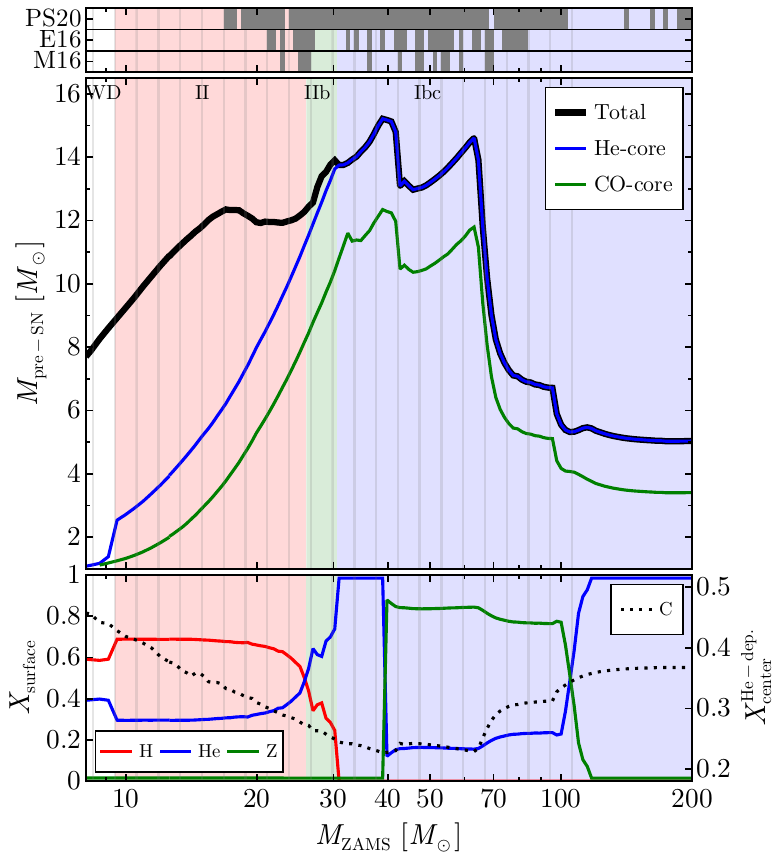}
    \caption{The features of single star models in Grid CC-S at the end of the run. The middle panel shows the final mass (black), He-core mass (blue) and CO-core mass (green). The bottom panel shows the surface composition (solid lines, left axis) and the central C abundance at core He-depletion (dotted line, right axis). The top panel highlights the models that are expected to implode (black background), for different explodability criteria. Background colors show the expected evolutionary outcome of a model in that mass range, either WD (white), \Type IIP/L (red), \Type IIb (green) or \Type Ibc supernova (blue).}
    \label{fig:singleSN}
\end{figure}

We report the properties at core-collapse for the single-star models in grid CC-S (Fig.\,\ref{fig:singleSN}), including the effect of the adoption of different explodability criteria. Single-stars of initial masses between $9.4\Msun$ and $26.0\Msun$ become progenitors of \Type IIP/L supernovae, followed by \Type IIb up to $30.5\Msun$ and \Type Ibc for higher masses. Using the explodability criterion from \Muller, we find that implosions affect progenitors with $M_\mathrm{ZAMS}\sim22-27\Msun$, as well as a few between $35-70\Msun$. Folding the explodability results with the IMF gives us that the relative proportion of \Type IIP/L, IIb and Ibc supernovae would be $80.2\%$, $5.1\%$ and $14.7\%$ respectively, with a \Type Ibc to \Type IIP/L ratio of $0.18$. For \Type IIP/L we find $\Mej = 7.6-10.7\Msun$ and $M_\mathrm{H,env}=2.8-7.0\Msun$, while for \Type IIb $\Mej=10.4-12.3\Msun$. 

\begin{figure}
    \centering
    \includegraphics[width=0.99\linewidth]{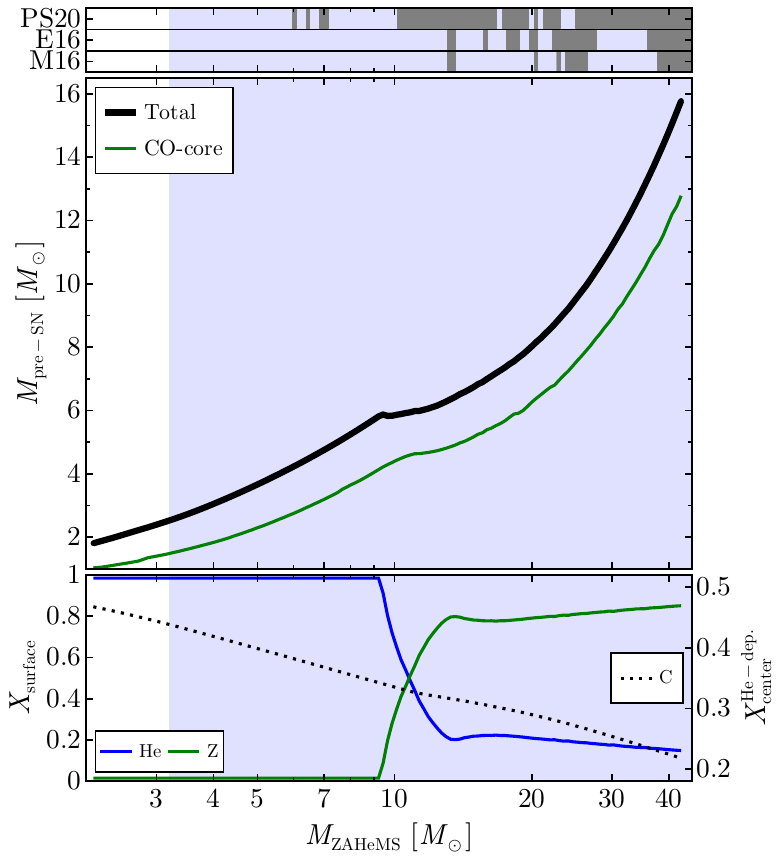}
    \caption{As Fig.\,\ref{fig:singleSN} but for single He-star models in Grid CC-He. }
    \label{fig:singleSN_HeS}
\end{figure}

The progenitors of \Type Ibc supernovae show varying masses distributed around two clearly distinct peaks. About $78\%$ of Ibc progenitors have $\Mej=11.3-13.3\Msun$ and the remaining $22\%$ around $3.8-6.5\Msun$. The second peak originates from the very massive stars at ZAMS ($M_\mathrm{i}>65\Msun$, cf. Fig.\,\ref{fig:singleSN}) that show severe mass-loss during the main-sequence. This is caused by the star showing strong He-enhancement early on in the main-sequence, which increases the mass-loss rate significantly and leads to the development of a smaller core. The number of BHs formed is about $8$ per 100 supernovae, with masses $M_\mathrm{BH}=10.3-14.2\Msun$.

When using the explodability criteria from \Ertl \ and \PS, the number of implosions increase, and the number of BHs formed significantly increases to $\sim20$ and $\sim72$ per 100 supernovae respectively. Figure\,\ref{fig:singleSN} shows the initial mass ranges that are mostly affected by implosions. \Type IIb and \Type Ibc supernovae are affected the most as many of their progenitors implode. For \Type IIP/L, the only progenitors that implode are the most massive. The relative fractions of supernovae change, generally by increasing \Type IIP/L at the expense of the other supernovae. The criterion from \Ertl \ implodes some of the higher-mass \Type Ibc progenitors, yielding  fractions of \Type IIP/L, IIb and Ibc of $85.2\%$, $4.7\%$ and $10.1\%$ respectively (and a \Type Ibc to \Type IIP/L ratio of $0.12$), while the criterion from \PS \ implodes almost all progenitors but those of \Type IIP/L with $M_\mathrm{ZAMS}\simle15\Msun$. 

While one cannot provide IMF-weighted distributions concerning single He-stars, we can nonetheless discuss the main trends that our models show in terms of their explodability (see Fig.\,\ref{fig:singleSN_HeS}). The criterion from \Muller \ predicts implosions for models with $M_\mathrm{pre-SN}\sim6.3\Msun$, $\sim 8.1\Msun$, $9-10\Msun$ and for $>13.1\Msun$. With the criterion from \Ertl, additional imploding models are found with final masses between $6.9\Msun$ and $10.4\Msun$, while that of \PS \ includes most models with $M_\mathrm{pre-SN}>5.9\Msun$ and a few low-mass ones around $4.2-4.7\Msun$.

\FloatBarrier

\section{Extrapolating Case BB RLOF}\label{appendix:CaseBB}

\begin{figure}
    \centering
    \includegraphics[width=\linewidth]{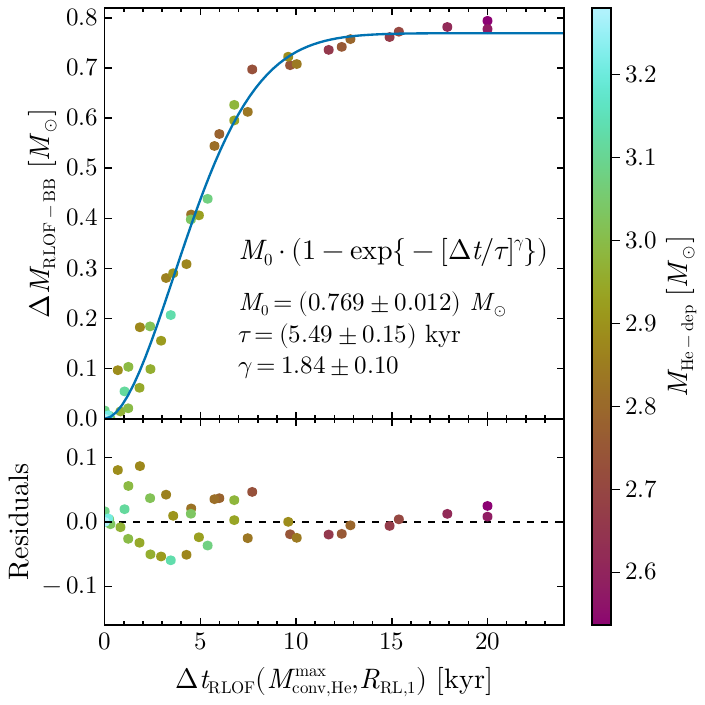}
    \caption{Fit of $\Delta M_\mathrm{RLOF-BB}$ from the models in \PII \ as a function of the time between the onset of \Case BB RLOF and core-collapse (see text). Each model is color-coded by the mass at core helium depletion $M_\mathrm{He-dep}$. The fitting function and parameters are shown in the top panel. The bottom panel shows the residuals.  }
    \label{fig:fit_Ibci}
\end{figure}

The models in \PII \ provide significant correlations between the information available before core-He depletion and the outcome of \Case BB RLOF (see Appendix\,B in \PII). For this, we need to know the maximum extent in mass of the convective core during core-He burning $M_\mathrm{conv,He}^\mathrm{max}$, the Roche-lobe radius at core-He depletion $R_\mathrm{RL,1}$ and a single-star model with the same $M_\mathrm{conv,He}^\mathrm{max}$. Therefore, we can use the single He star model and $R_\mathrm{RL,1}$ to derive the time before core-collapse where this He-star would fill its Roche-lobe $\Delta t_\mathrm{RLOF-BB}$. We calculate this value for the models shown in \PII \ (using the He-stars in the same work) and plot their $\Delta M_\mathrm{RLOF-BB}$ against $\Delta t_\mathrm{RLOF}$ (Fig.\,\ref{fig:fit_Ibci}). 

We constructed a fitting function that provides a good approximation to the behavior seen in the models,
\begin{equation}
    \Delta M_\mathrm{RLOF-BB}(\Delta t_\mathrm{RLOF-BB}) = M_0\left(1-\mathrm{e}^{-\left[{\Delta t_\mathrm{RLOF-BB}}/{\tau}\right]^\gamma}\right),
\end{equation}
where $M_0$ gives the maximum mass-loss, $\tau$ a characteristic timescale and $\gamma$ a power-exponent. With our fit, we obtain a relatively good agreement with the models, with relatively limited residuals of up to $0.1\Msun$ in the low $\Delta t_\mathrm{RLOF-BB}$ regime. 

\FloatBarrier

\section{Example of the effects of binary interaction}\label{sec:SNprog:Binary_Grid}

\begin{figure}
    \centering
    \includegraphics[width=1\linewidth]{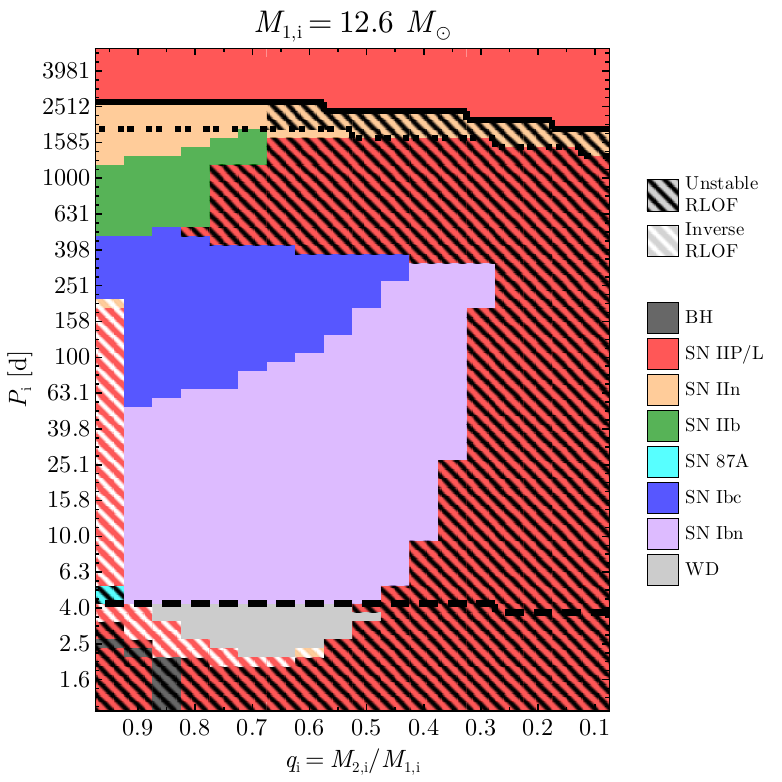}
    \caption{Expected evolutionary outcome for the primary star, or merger product, for each model in the binary grid with $M_\mathrm{1,i}=12.6\Msun$ shown in the $\log P_\mathrm{i}-\qi$ diagram. The explodability criterion used is that of \Muller  \ and the mergers are given by the ``Hardcoded'' criteria. Models that undergo unstable RLOF are shown with hatches, and we differentiate between the case in which the donor is the primary star (black) or the secondary star (white).  Different colors represent different evolutionary outcomes (see legend).
    The lines in the plots separate \Case A and \Case B systems (dashed), \Case B and \Case C systems (dotted), and systems that undergo RLOF and those that do not (solid). 
    }
    \label{fig:logPq_1.10}
\end{figure}

Binary evolution influences the evolution of the primary star by removing its hydrogen-rich envelope prior to the end of its life. Conversely, this process can help produce more massive progenitors via accretion or mergers. In Fig.\,\ref{fig:logPq_1.10} we illustrate how varying initial orbital configurations lead to distinct levels of stripping, consequently resulting in diverse supernova types for a primary star of $M_\mathrm{1,i}=12.6\Msun$ (which, if isolated, would evolve to become a \Type IIP/L supernova).

The tightest binaries undergo mass-transfer during the main-sequence (\Case A RLOF, \citealt{Pols94_CaseA, Sen2022_modelsCaseA}). Systems with $P_\mathrm{i}\simle2\days$ mostly develop into contact systems that undergo L2-outflows and thus merge \citep{Menon21_contactbinaries}. Some of these mergers are so massive that they are then expected to produce BHs instead of exploding as a supernova. For those systems with $P_\mathrm{i}\sim2-4\days$, models undergo stable \Case A only for $0.50\simle\qi\simle0.85$, which is followed by subsequent phases of RLOF and wind mass-loss to strip the star so severely that it will turn into a WD. For lower $\qi$, \case A turns unstable, while for higher $\qi$ the companion stars evolve quickly enough to the point that they initiate inverse mass-transfer, which is assumed to end up in a merger (see Sect.\,\ref{sec:methods:binary_physics}). Such mergers typically produce \Type IIP/L supernovae, but in some cases also give rise to \Type IIb supernovae (due to the small amount of H-rich envelope that was left), and in one model this occurs so late that core-collapse may soon follow, leading to a \Type IIn supernova. 

For wider binaries, with  $4\days\simle P_\mathrm{i}\simle1\,700\days$, the first phase of mass transfer occurs after the end of the main-sequence and before the beginning of core He burning (\Case B RLOF). Mass-transfer efficiently removes a significant part of the H-rich envelope of the primary star, to the point that for $P_\mathrm{i}<400\days$ the subsequent winds are able to remove the remaining portion of the H-rich envelope, producing \Type Ibc supernova progenitors. Additionally, there is a significant parameter space for stripped He-star progenitors to expand once again and undergo \Case BB RLOF, becoming progenitors of \Type Ibn supernovae. For wider orbits, mass-transfer and wind mass-loss are not strong enough to remove the whole envelope, and the primary will explode as a \Type IIb supernova.

For $1\,000\simle P_\mathrm{i}\simle2\,500\days$, the orbit is wide enough that RLOF can occur after core-He depletion (\Case C mass transfer), which will lead to a \Type IIn supernova (see Sect.\,\ref{sec:methods:sn_type}). This also applies to models that first underwent \Case B RLOF but retained massive envelopes (\PI). For even wider binaries ($P_\mathrm{i}\simgr2\,500\days$), no mass-transfer occurs and therefore the primary star evolves as a single-star, which will then explode as a \Type IIP/L supernova.  This behavior is similar for different masses, although the supernova types differ.

\FloatBarrier

\section{Comparison to other theoretical predictions}\label{sec:disussion:other_works}

Works like \cite{Eldridge08_massivebinaries_sntypes} and \cite{Souropanis25_SESNe_Metallicity_POSYDON} discuss the relative population of \Type Ibc to \Type IIP/L supernovae from theoretical models accounting for single-star and binary evolution across different metallicities. In \cite{Eldridge08_massivebinaries_sntypes}, similar-metallicity models to our own result in \Type Ibc to \Type IIP/L ratios of $\sim0.3$, compared to our value of $0.48$, as they neglect rotation, which can boost the appearance of H-free supernovae \citep[e.g.,][]{MeynetMaeder2005_rotation_WR_and_SN}. The models from \cite{Souropanis25_SESNe_Metallicity_POSYDON}, also predict a similar ratio, although their models use a smaller binary fraction ($f_B=60\%)$, which also reduces their number of \Type Ibc.  

We also compare the results from \cite{Zapartas19_howmany_Hrich_Sne_from_binaries} with a similar $f_B$ to our own and find a qualitatively similar ratio of \Type Ibc to \Type IIP/L ($\sim0.39$). This work also extensively focused on the progenitor channels for \Type IIP/L supernovae, where they find that only $35\%$ of \Type IIP/L progenitors come from single and effectively-single stars, $20\%$ from donors and accretors, while the remaining $45\%$ from mergers. This is in line with our models,  where we obtain respectively $37\%$, $23\%$ and $40\%$. 

The recent work by \cite{Gilkis25_CCSNe_WR} offers a more systematic comparison of different supernova types for the models at the same metallicity. They find about $67\%$ \Type IIP/L supernovae ($M_\mathrm{H,env}>0.03$), $18.3\%$ for  \Type IIb (where $0.030<M_\mathrm{H,env}/M_\odot<0.001$) and $14.6\%$ of \Type Ibc supernovae. Their models produce significantly more \Type IIb supernovae than our models due to the different wind recipe adopted for lower-mass partially-stripped stars \citep{Vink_WR}. However, their number of \Type Ibc supernovae are in tension with those in \cite{Graur2017_LOSS}, which our models better replicate, while they are within $5\%$ of the estimate from \cite{Ma25_SNpops40Mpc}. The number of BHs they produce is also a factor four higher than our models (assuming the criterion from \Muller, which they also adopt), which can be explained by the generally higher masses that their stars show due to the weaker winds. Finally, the authors also observe an enhancement in mass-loss rates for stripped stars as they advance through their later evolutionary phases, attributed to the formation of optically-thick winds. This aspect is absent in our models, due to the divergent wind schemes adopted.

\Type IIn supernovae have been investigated in  \cite{Schneider25_accretorsmergerSNe_IIP_87A_IIn} using detailed single-star models to mimic the evolution of accretors and mergers in binary systems until core-collapse \citep{Schneider2024_preSNevo_mergers_secondaries}. They assume that \Type IIn supernovae are formed when the star explodes while located above the Humphrey-Davidson limit \citep{HumpreysDavidson1979} and cooler than the sDor instability strip \citep{Smith2004_sDor_LBV}, under the assumption that a star in this region would show significant mass-loss prior to the explosion. While we cannot assess the evolution of merger models in the HRD, we have some secondary stars that do reach core-collapse while located in this region (see Fig.5 in \citealt{Jin25_BonnGal}) which contribute to about $0.3\%$ of all core-collapse supernovae. Even if they were to become \Type IIn supernovae, our overall results would not be significantly affected.

The only study released so far with regards to the population of \Type Ibn supernovae is from \cite{Ko25_PopSynth_Ibn}, where the authors use a population synthesis code \citep{Kinugawa14_popsynth1_popIII, Kinugawa16_popsynth2_popIII, Kinugawa24_popsynth3_SNe_binaries} to produce the population of these transients through mass-transfer with a main-sequence star or compact-object. They find that their \Type Ibn supernovae account for $1-3\%$ of all supernovae, with $>85\%$ of them coming from He-stars transferring mass to a secondary main-sequence star (see their Fig.4). Their results are not only compatible with our results, but also justify our assumption of exclusively studying the primary-star progenitors for \Type Ibn supernovae.

\FloatBarrier

\section{Ratios}

\begin{figure}
        \centering
    \includegraphics[width=\linewidth]{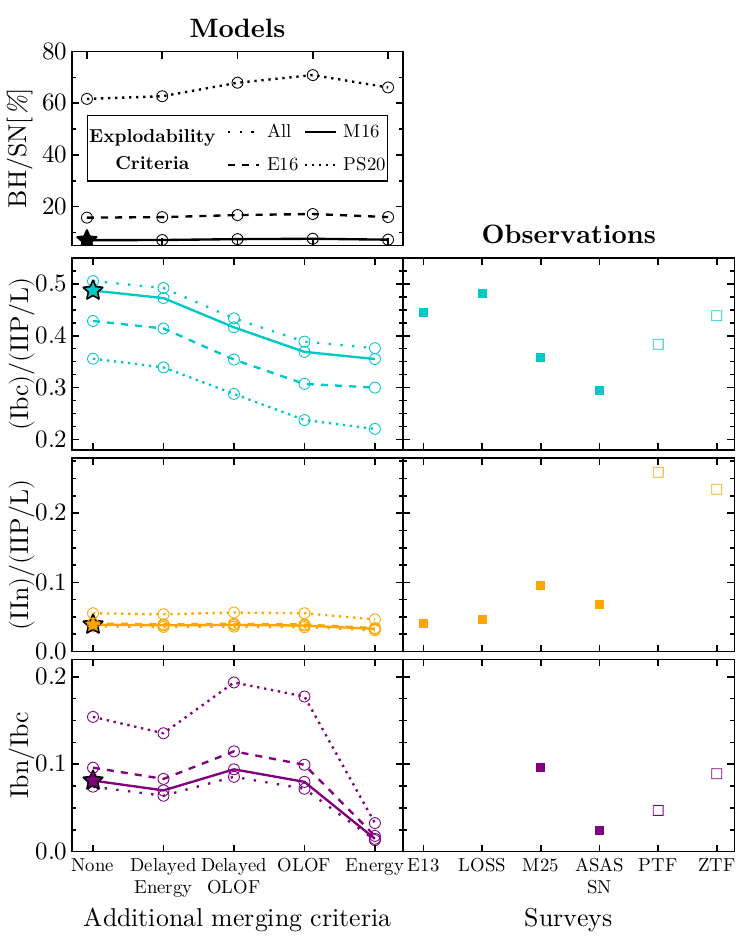}
    \caption{As Fig.\,\ref{fig:fractions_different_criteria} but showing ratios between transients or BHs formed as a function of different merger criteria (see Fig.\,\ref{fig:fractions_different_criteria}), sorted from left to right by those that produce less to more mergers. We show the ratio of BHs to supernovae (top left, in percent), \Type Ibc to \Type IIP/L supernovae (bottom left), \Type IIn to \Type IIP/L (top right) and \Type Ibn to \Type Ibc (bottom right).  Different dashing show the same while assuming different explodability criteria. The left sub-panels show the predictions from the models, while on the right sub-panels those derived from observations.}
    \label{fig:BHs_different_criteria}
\end{figure}

Figure\,\ref{fig:BHs_different_criteria} shows complementary information to Fig\,\ref{fig:fractions_different_criteria} including the number of BHs and the ratios of different supernova types.  

\end{appendix}

\end{document}